\documentclass[aps,pre,reprint,twocolumn]{revtex4-1}
\usepackage[latin9]{inputenc}
\usepackage{array}
\usepackage{float}
\usepackage{multirow}
\usepackage{amsmath}
\usepackage{amstext}
\usepackage{amssymb}
\usepackage{graphicx}
\usepackage[unicode=true,pdfusetitle,
 bookmarks=true,bookmarksnumbered=false,bookmarksopen=false,
 breaklinks=false,pdfborder={0 0 1},backref=false,colorlinks=false]
 {hyperref}
\usepackage{breakurl}
\usepackage{ulem}

\makeatletter

\usepackage{color}
\usepackage{epsfig}
\usepackage{graphics}
\usepackage{epsfig}
\usepackage{tikz}
\usetikzlibrary{shapes.geometric}
\usepackage{epstopdf}
\usepackage{epsfig}

\begin{document}

\title{Coupled elastic membranes
model for quantum heat transport in semiconductor nanowires}

\author{Julian A Lawn and Daniel S Kosov}

\affiliation{College of Science and Engineering, James Cook University,	Townsville, QLD, 4811, Australia}

\begin{abstract}
	Presented here is a nanowire model, consisting of coupled elastic membranes with the purpose of investigating thermal transport in quasi-one-dimensional quantum systems. The vibrations of each elastic membrane are quantized and the flow of the vibrational energy between adjacent membranes is allowed. The ends of the nanowire are attached to thermal baths held at different temperatures. We derived quantum master equation for energy flow across the nanowire and obtained thermal  currents and other key observables. We study the effects of a disordered boundary on the thermal current by randomizing the membrane radii. We evaluate the model as a nanowire analogue as well as study the effects of a disordered boundary on thermal conductivity.  The calculations show that the membrane lattice model demonstrates diameter phonon confinement and a severe reduction in thermal conductivity due to surface roughness which is characteristic of semiconductor nanowires. The  surface roughness also produces a length dependence of the thermal conductivity of the form $\kappa=\alpha L^{\beta}$, with $\beta$ dependent on disorder characteristics, in the otherwise ballistic regime. Finally, the parameters of the model are fitted to available experimental data for silicon nanowires and the results of the calculations are assessed against the experimental data.
\end{abstract}
\maketitle

\section{Introduction}

Thermal transport in bulk materials is generally well described by Fourier's law $J=-\kappa\,\triangledown\,T$, where the thermal current is driven by the local temperature gradient proportional to the thermal conductivity, $\kappa$. The thermal conductivity was considered to be an intrinsic property of material, independent of the geometry the material. However it was shown that for a one dimensional lattice of coupled harmonic oscillators that thermal energy was transported ballistically  \cite{Rieder1967}, like a wave, which is sometimes referred to as ``second sound''  \cite{Peshkov1944,Guyer1966}. Since then it has been shown for low dimensional systems that the thermal conductivity diverged with system length $L$ such that $\kappa\sim L^{\beta}$, such as disordered harmonic\cite{Casher1971,Hu1998} and anharmonic chains \cite{Hu1998,Lepri1997}, truncated Toda lattices\cite{Hatano1999} and Fermi-Ulam-Pasta chains\cite{lepri1998,Narayan2002}.

Semiconductor nanowires have  been of interest due to the reduced thermal conductivity that they exhibit in comparison to bulk materials \cite{Li2003}. This presented a theoretical challenge and an opportunity for technological application, particularly for thermo-electric devices \cite{Humphrey2005}. It was shown that the small diameter of silicon nanowires compared to phonon wavelengths reduced the thermal conductivity \cite{Li2003}. Due to large surface to volume ratio of nanowires, thermal conductivity is also significantly affected by the geometry of the surface. In experimental samples where the surfaces is disordered due to etching or corrugation, the thermal conductivity is reduced towards the amorphous limit \cite{Hochbaum2008}. The combination of the reduced dimensions and roughness effects led to numerous theoretical works describing the phenomena \cite{Mingo2003,Liang2006,Moore2008,Donadio2009,Martin2009,Kosevich2009,Luisier2011,Blanc2013}. The reduction of the thermal conductivity due to surface roughness can reduce the thermal conductivity below Casimir's classical limit where boundary effects dominate thermal characteristics \cite{Carrete2011,Sadhu2011,Blanc2013}. It was also found, both theoretically \cite{Sadhu2011} and experimentally \cite{Lim2012}, that the correlation length of the surface roughness plays a significant role, with shorter correlation lengths being key to reducing the thermal conductivity. Additionally, computer simulations using classical molecular dynamics  suggested that nanowire lengths below the phonon mean free path limit phonon-phonon interactions, leading to super-diffusive behaviour and length dependence on thermal conductivity \cite{Yang2010} where the thermal conductivity  written as a function of the nanowire length $\kappa=\alpha L^{\beta}$ was linearly dependent ($\beta=1)$ on the system length, up to around 60nm; beyond which the exponent $\beta$ reduces towards Fourier like behaviour. Recently semi-ballistic phonon transport was confirmed experimentally for silicon nanowires at temperatures around $4$K \cite{Maire_2017}.

The model developed here, constructs a nanowire analogue out of a series of coupled elastic membranes which each have their own (local) vibrational spectrum which depends on the membrane radius. By choosing the size of each individual membrane the surface roughness directly influences the local vibrational spectra of the nanowire. The system is coupled weakly at either end to thermal reservoirs held at different temperatures to drive a thermal flux resulting in a combined system-environment Hamiltonian. In a quantum master equation approach the nanowire is treated as an open quantum system. The application of the sequence of approximations, the Born approximation (keeping terms up to second order for nanowire-environment coupling in the Liouville equation for the reduced density matrix), the Markov approximation (assumption that the correlation functions of the electrodes decay on a time scale much faster than tunneling events) and the rotating wave approximation, leads  to a Lindblad master equation. The steady state observables, such as thermal conductivity, are derived analytically using this master equation.

The outline of the paper is as follows. Section II describes the general theory, including the model Hamiltonian, derivation of the quantum master equation and expressions for key physical observables.  In section III we applied the theory to study heat transport through  model semiconductor nanowires with different types of static surface disorder. Section IV summarizes the main results of the paper.

\section{Theory}

\subsection{Physical model and Hamiltonian}

The undoped semiconductor nanowire conducts heat by phonons and contribution
from  electron and electron-hole degrees of freedom is negligible.  To develop a physical model for heat transport, we "slice" the nanowire into thin discs. The width of the disk is assumed to be much smaller than radius of the nanowire and we assume that the disc vibration spectrum can be approximated by the eigen-frequencies of a vibrating thin elastic membrane.

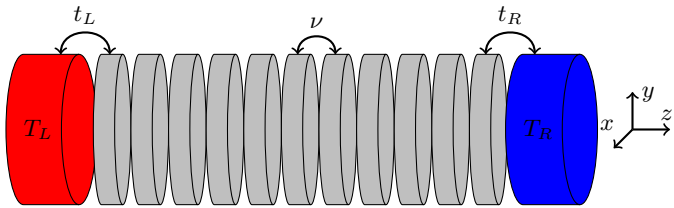
\begin{figure}[H]
	\centering{}\begin{tikzpicture} \node [cylinder, fill=red, rotate=0, draw, minimum height=1cm, minimum width=2cm] (TL) at (-0.4,0) {$T_L$};
	\node [cylinder, fill=lightgray, rotate=0, draw, minimum height=0.5cm, minimum width=2cm] (a1)  at (0.5,0) {};
	\node [cylinder, fill=lightgray, rotate=0, draw, minimum height=0.5cm, minimum width=2cm] (a2)  at (2*0.5,0) {};
	\node [cylinder, fill=lightgray, rotate=0, draw, minimum height=0.5cm, minimum width=2cm] (a3)  at (3*0.5,0) {};
	\node [cylinder, fill=lightgray, rotate=0, draw, minimum height=0.5cm, minimum width=2cm] (a4)  at (4*0.5,0) {};
	\node [cylinder, fill=lightgray, rotate=0, draw, minimum height=0.5cm, minimum width=2cm] (a5)  at (5*0.5,0) {};
	\node [cylinder, fill=lightgray, rotate=0, draw, minimum height=0.5cm, minimum width=2cm] (a6)  at (6*0.5,0) {};
	\node [cylinder, fill=lightgray, rotate=0, draw, minimum height=0.5cm, minimum width=2cm] (a7)  at (7*0.5,0) {};
	\node [cylinder, fill=lightgray, rotate=0, draw, minimum height=0.5cm, minimum width=2cm] (a8)  at (8*0.5,0) {};
	\node [cylinder, fill=lightgray, rotate=0, draw, minimum height=0.5cm, minimum width=2cm] (a9)  at (9*0.5,0) {};
	\node [cylinder, fill=lightgray, rotate=0, draw, minimum height=0.5cm, minimum width=2cm] (a10)  at (10*0.5,0) {};
	\node [cylinder, fill=lightgray, rotate=0, draw, minimum height=0.5cm, minimum width=2cm] (a11)  at (11*0.5,0) {};
	\node [cylinder, fill=blue, rotate=0, draw, minimum height=1cm, minimum width=2cm] (TR) at (12.5*0.5,0) {$T_R$};
	\draw [<->,thick,out=90,in=90,looseness=1.6] (6*0.5+0.05,1) to node[above]{$\nu$}  (7*0.5+0.05,1);
	\draw [<->,thick,out=90,in=90,looseness=1.5] (-0.1,1) to node[above]{$t_{L}$}  (0.55,1);
	\draw [<->,thick,out=90,in=90,looseness=1.5] (11*0.5+0.05,1) to node[above]{$t_{R}$} (6.2,1);
	
	\draw [->,thick] (7.5,0) to node[above right]{$z$} (8,0);
	\draw [->,thick] (7.5,0) to node[above right]{$y$} (7.5,0.5);
	\draw [->,thick] (7.5,0) to node[above left]{$x$} (7.25,-0.25);

	\end{tikzpicture} 
	
	\caption{Sketch of the coupled elastic membranes model of quantum wire.}
	\label{sketch} 
\end{figure}

The vibration of a thin membrane is, for small displacements, modelled
well by the two-dimensional wave equation 
\begin{equation}
\frac{\partial^{2}u}{\partial t^{2}}=c^{2}\left(\frac{\partial^{2}u}{\partial x^{2}}+\frac{\partial^{2}u}{\partial y^{2}}\right),
\end{equation}
Here $u(x,y,t)$ is the displacement of the membrane from its equilibrium
position, which assumed to be zero on the surface of nanowire. Here
$c$ is a property of the particular nanowire material and its value
will be determined later. We use cylindrical coordinates and also
assume that the membrane vibrations have radial symmetry, that means
$u(r,\phi,t)=u(r,t)$. The wave equation becomes 
\begin{equation}
\frac{\partial^{2}u}{\partial t^{2}}=c^{2}\left(\frac{\partial^{2}u}{\partial r^{2}}+\frac{1}{r}\frac{\partial u}{\partial r}\right).
\end{equation}
This differential equation is solved by the separation of variables
and the solution has the following time dependence 
\begin{equation}
u(r,t)\sim e^{i\lambda ct}f(r).
\end{equation}
Therefore, the period of vibration is $2\pi/\lambda c$, where $c$ is a parameter which has dimension of velocity. Parameter
$\lambda$ is given through zeros of a Bessel function of the first kind \cite{Asmar2005}
\begin{equation}
J_{0}(\lambda_{k}R)=0,\;\;\;\;k=1,2,3,...\label{eq:dispersion}
\end{equation}
where $R$ is the radius of the membrane. The corresponding quantized energy spectrum  of elastic membrane is $E_{k}=h\lambda_{k}c/2\pi=\hbar\lambda_{k}c$.

Therefore, the quantized Hamiltonian for the individual elastic membrane
is 
\begin{equation}
\sum_{k}E_{k}(b_{k}^{\dagger}b_{k}+\frac{1}{2}),\label{eq:Membrane H}
\end{equation}
where $b_{k}^\dagger$ and $b_{k}$ are bosonic creation and annihilation
operators satisfying the standard commutation relations 
\begin{equation}
[b_{k},b_{k'}^{\dag}]=\delta_{kk'},\;\;\;[b_{k},b_{k'}]=[b_{k}^{\dag},b_{k'}^{\dag}]=0.
\end{equation}

We introduce coupling
between membranes to enable vibrational energy transfer between them.
The Hamiltonian of the quantum wire becomes

\begin{align}
H_{W}=&\sum_{\alpha}^{N}\sum_{k}E_{k}b_{\alpha k}^{\dagger}b_{\alpha k} \nonumber\\
&+\sum_{\alpha}^{N-1}\sum_{kk'}v_{kk'}\left[b_{\alpha k}^{\dagger}b_{\alpha+1k'}+b_{\alpha+1k}^{\dagger}b_{\alpha k'}\right],
\end{align}
where $\alpha$ is the membrane index and $N$ is the number of
membranes that comprise the wire. We omitted the zero point energy part
of this Hamiltonian, since it does not contribute to any quantities
we compute. The amplitude for inter-membrane coupling  is taken in the form 
\begin{equation}
v_{kk'}=\frac{v_{0}}{\left|E_{\alpha k}-E_{\alpha+1k'}\right|/q+1}.\label{eq:hopping}
\end{equation}
The physical reasoning behind this choice of the interaction is the following. It describes the transfer of energy between the vibrations of the nearest-neighbour membranes, it is maximal when the membrane vibrational energy levels are in resonance and decays as the energy gap between the vibrational states of interest grows. Value $v_0$ gives the strength of inter-membrane coupling and parameter $q$ suppresses (for large $q$) or amplifies (for small $q$) the resonance-dominated energy transfer. The wire is attached to two thermal baths --- macroscopically large ideal gases of phonons with spectra $\epsilon_{l}$ and $\epsilon_{r}$  held at different temperatures $T_{L}$ and $T_{R}$, respectively. The Hamiltonians for the thermal baths are \begin{equation} H_{L}=\sum_{l}\epsilon_{l}b_{l}^{\dag}b_{l},\;\;\;\;\;H_{R}=\sum_{r}\epsilon_{r}b_{r}^{\dag}b_{r}. \end{equation} Phonon-exchange interaction between the wire and baths is taken in the energy transfer form

\begin{equation}
V=t_{L}\sum_{lk}\left(b_{l}^{\dag}b_{1k}+b_{1k}^{\dag}b_{l}\right)+t_{R}\sum_{rk}\left(b_{r}^{\dag}b_{Nk}+b_{Nk}^{\dag}b_{r}\right).
\end{equation}

Here the first term describes the vibration quanta transfer between
the left bath and the first membrane with amplitude $t_{L}$ and the
second term describes the vibration quanta transfer between the last
membrane and the right bath with amplitude $t_{R}$.

\subsection{Canonical transformation from local membrane vibration to wire normal modes}

The wire is described by the Hamiltonian which is in bosonic quadratic
form:

\begin{equation}
H_{W}=\sum_{\alpha k}\sum_{\alpha'k'}h_{\alpha k,\alpha'k'}b_{\alpha k}^{\dagger}b_{\alpha'k'},\label{eq:NW Hamiltonian}
\end{equation}

where the $h_{\alpha k,\alpha'k'}$ are the matrix elements of the block-tridiagonal Hamiltonian matrix

\begin{centering}
\begin{equation}
h=\left[\begin{array}{ccccc}
h_{\alpha k} & v_{kk'} &  &  & 0\\
v_{kk'} & \ddots & v_{kk'}\\
& v_{kk'} & h_{\alpha k} & v_{kk'}\\
&  & v_{kk'} & \ddots & v_{kk'}\\
0 &  &  & v_{kk'} & h_{\alpha k}
\end{array}\right]
\end{equation}
\end{centering}

where

\begin{equation}
h_{\alpha k} = \left[\begin{array}{ccc}
E_{\alpha1}\\
& \ddots\\
&  & E_{\alpha K_{\alpha}}
\end{array}\right]
\end{equation}

is the primary block diagonal consisting
of the diagonal matricies of local vibrational energies $E_{\alpha k}$,
and the upper and lower block diagonals containing the inter-membrane
coupling elements $v_{kk'}$. This Hamiltonian is diagonalized by
the canonical transformation  from initial localized membrane vibrations
$b_{\alpha k}^{\dag}$ to global nanowire normal modes $B_{s}^{\dag}$

\begin{equation}
B_{s}^{\dag}=\sum_{\alpha k}U_{\alpha k,s}b_{\alpha k}^{\dag}\;\;\;\;B_{s}=\sum_{\alpha k}U_{\alpha k,s}b_{\alpha k}.
\end{equation}

The transformation is unitary in order to preserve the bosonic commutation
relations

\begin{equation}
\sum_{\alpha k}U_{\alpha k,s}U_{\alpha k,s'}=\delta_{ss'},\;\; \sum_{s}U_{\alpha k,s}U_{\alpha'k',s}=\delta_{kk'}\delta_{\alpha\alpha'}
\end{equation}

If we require that, the transformation matrix satisfies the eigenvalue
problem

\begin{equation}
\sum_{\alpha'k'}h_{\alpha k,\alpha'k'}U_{\ \alpha'k',s}=\epsilon_{s}U_{\alpha k,s},
\end{equation}

then the Hamiltonian becomes diagonal in normal mode creation and annihilation
operators

\begin{equation}
H_{W}=\sum_{s}\epsilon_{s}B_{s}^{\dag}B_{s}.
\end{equation}

The coupling of the wire to the thermal bath is also transformed to
normal mode operators

\begin{equation}
V=\sum_{ls}\nu_{Ls}\left(b_{l}^{\dag}B_{s}+B_{s}^{\dag}b_{l}\right)+\sum_{rks}\nu_{Rs}\left(b_{r}^{\dag}B_{s}+B_{s}^{\dag}b_{r}\right),
\label{vsb}
\end{equation}

where we introduced the amplitudes for energy transfer between phonons in the thermal bath and nanowire normal modes

\begin{equation}
\nu_{Ls}=t_{L}\sum_{k}U_{1k,s}\;\;\;\;\nu_{Rs}=t_{R}\sum_{k}U_{Nk,s}.
\end{equation}

\subsection{Quantum master equation for nanowire density matrix and observables}

Using the Born-Markov and rotating wave approximations (details of the
derivation are given in appendix A) we get a Lindblad type master equation
for the nanowire density matrix $\rho(t)$

\begin{eqnarray}
i\hbar\dot{\rho}(t) = & \left[\sum_{s}\tilde{\epsilon}_{s}B_{s}^{\dagger}B_{s},\rho(t)\right] \\
&-i\sum_{s, \mu=L,R}\Gamma_{\mu}(s)\left[\frac{1}{2}(1+2n_{\mu}(s))\{B_{s}^{\dag}B_{s},\rho(t)\} \right.  \nonumber \\
& \left. -(1+n_{\mu}(s))B_{s}\rho B_{s}^{\dag}-n_{\mu}(s)B_{s}^{\dag}\rho B_{s}+n_{\mu}(s)\rho\right] \nonumber 
\end{eqnarray}

The energies of the nanowire normal modes are shifted by the coupling to
the left and right baths

\begin{equation}
\tilde{\epsilon}_{s}=\epsilon_{s}+\Delta_{L}(s)+\Delta_{R}(s)
\end{equation}

by the Lamb shifts (real parts of bath self-energies)

\begin{eqnarray}
\Delta_{L}(s) & =[\nu_{Ls}]^{2}\sum_{l}\frac{\epsilon_{s}-\epsilon_{l}}{(\epsilon_{l}-\epsilon_{s})^{2}+v^{2}}\nonumber \\
\Delta_{R}(s) & =[\nu_{Rs}]^{2}\sum_{r}\frac{\epsilon_{s}-\epsilon_{r}}{(\epsilon_{r}-\epsilon_{s})^{2}+v^{2}}.\label{eq:E renormalization}
\end{eqnarray}

The dissipative part of the density matrix time evolution is governed
by the normal mode energy level broadening function (imaginary parts
of the baths self-energies)

\begin{equation}
\Gamma_{L}(s)=2\pi[\nu_{Ls}]^{2}\rho_{L}(s)\;\;\;\;\Gamma_{R}(s)=2\pi[\nu_{Rs}]^{2}\rho_{R}(s).\label{eq:gamma}
\end{equation}

Let us now demonstrate how to use this master equation to compute
observables of interest. The average value of an arbitrary operator $O$
at time $t$ is 
\begin{equation}
\langle O \rangle_{t} = \text{Tr}[O\rho(t)].
\end{equation}

Differentiating with respect to $t$ gives the equation of motion
for a time-dependent expectation value of operator $O$: 

\begin{multline}
i\hbar\frac{d}{dt}\langle O\rangle_{t}=\sum_{s}\tilde{\epsilon}_{s}\langle[O,B_{s}^{\dag}B_{s}]\rangle_{t} \\
-i\sum_{s,\mu=L,R}\Gamma_{\mu}(s)\Big[\frac{1}{2}(1+2n_{\mu}(s))\langle[B_{s}^{\dag}B_{s},O]\rangle_{t} \\
 -n_{\mu}(s)\langle[B_{s},O]B_{s}^{\dag}\rangle_{t}-(1+n_{\mu}(s))\langle[B_{s}^{\dag},O]B_{s}\rangle_{t}\Big]
\label{eofm}
\end{multline}

Using (\ref{eofm}) we compute various nonequilibrium quantities to
characterize properties of the nanowire. We begin with the number
of vibrational quanta populating the normal mode $s$. Substituting
the corresponding operator into (\ref{eofm}) we get $n_{s}=B_{s}^{\dag}B_{s}$

\begin{align}
i\hbar\frac{d}{dt}\langle n_{s}\rangle_{t}= i\sum_{\mu=L,R}&\Gamma_{\mu}(s)\Big[n_{\mu}(s)\langle[B_{s},n_{s}]B_{s}^{\dag}\rangle_{t}\nonumber\\
&+(1+n_{\mu}(s))\langle[B_{s}^{\dag},n_{s}]B_{s}\rangle_{t}\Big],
\end{align}

which gives

\begin{equation}
\frac{d}{dt}\langle n_{s}\rangle_{t}=\sum_{\mu=L,R}\frac{\Gamma_{\mu}(s)}{\hbar}\left[n_{\mu}(s)-\langle n_{s}\rangle_{t}\right].
\end{equation}

This differential equation has the following solution:

\begin{align}
\langle n_{s}\rangle_{t} =& \frac{\Gamma_{L}(s)n_{L}(s)+\Gamma_{R}(s)n_{R}(s)}{\Gamma_{L}(s)+\Gamma_{R}(s)}
\nonumber\\
&+\exp\left(-\frac{\Gamma_{L}(s)+\Gamma_{R}(s)}{\hbar}t\right) \Big[\langle n_{s}\rangle_{t=0}\nonumber\\
&\hskip 5em -\frac{\Gamma_{L}(s)n_{L}(s) +\Gamma_{R}(s)n_{R}(s)}{\Gamma_{L}(s)+\Gamma_{R}(s)} \Big] 
\end{align}

If we tend $t$ to infinity, then the nanowire reaches a nonequilibrium steady state regime and the 
populations of the nanowire normal modes become stationary
\begin{eqnarray}
\langle n_{s}\rangle=\frac{\Gamma_{L}(s)n_{L}(s)+\Gamma_{R}(s)n_{R}(s)}{\Gamma_{L}(s)+\Gamma_{R}(s)}.\label{population}
\end{eqnarray}

Having computed the nonequilibrium population of nanowire normal modes
(\ref{population}) we can easily compute the amount of vibrational
energy stored in the  nanowire
\begin{equation}
\langle H_{W}\rangle_{t}= \sum_{s}\epsilon_{s}\frac{\Gamma_{L}(s)n_{L}(s)+\Gamma_{R}(s)n_{R}(s)}{\Gamma_{L}(s)+\Gamma_{R}(s)}.
\end{equation}
We define the heat current  using the continuity equation for the vibrational
energy flow. The energy conservation gives 
\begin{equation}
\frac{d}{dt}\langle H_{W}\rangle_{t}=J_{L}(t)+J_{R}(t),\label{continuity}
\end{equation}
where $J_{L} (J_{R})$ is the energy flowing into the system
from the left (right) heat baths. The rate of change of the vibrational
energy stored in nanowire is 
\begin{align}
\nonumber
\frac{d}{dt}\langle H_{W}\rangle_{t} =\sum_{s}\epsilon_{s}&\left\{\frac{\Gamma_{L}(\epsilon_{s})}{\hbar}[n_{L}(\epsilon_{s})-\langle n_{s}\rangle_{t}]\right.\\ &\left.+\frac{\Gamma_{R}(\epsilon_{s})}{\hbar}[n_{R}(\epsilon_{s})-\langle n_{s}\rangle_{t}]\right\} 
\label{conservation}
\end{align}
Comparing (\ref{conservation}) with the right hand side of the continuity
equation (\ref{continuity}), we identify the heat currents for the
energy flowing into the wire from the left bath 
\begin{equation}
J_{L}(t)=\sum_{s}\epsilon_{s}\frac{\Gamma_{L}(\epsilon_{s})}{\hbar}[n_{L}(\epsilon_{s})-\langle n_{s}\rangle_{t}],
\end{equation}
and from the right bath 
\begin{equation}
J_{R}(t)=\sum_{s}\epsilon_{s}\frac{\Gamma_{R}(\epsilon_{s})}{\hbar}[n_{R}(\epsilon_{s})-\langle n_{s}\rangle_{t}].
\end{equation}
In the steady state regime, the current becomes time-independent  and it is given by
\begin{equation}
J_{L}=\frac{1}{\hbar}\sum_{s}\epsilon_{s}\frac{\Gamma_{L}(\epsilon_{s})\Gamma_{R}(\epsilon_{s})}{\Gamma_{L}(\epsilon_{s})+\Gamma_{R}(\epsilon_{s})}[n_{L}(\epsilon_{s})-n_{R}(\epsilon_{s})],
\end{equation}
and
\begin{equation}
J_{R}=\frac{1}{\hbar}\sum_{s}\epsilon_{s}\frac{\Gamma_{L}(\epsilon_{s})\Gamma_{R}(\epsilon_{s})}{\Gamma_{L}(\epsilon_{s})+\Gamma_{R}(\epsilon_{s})}[n_{R}(\epsilon_{s})-n_{L}(\epsilon_{s})].
\end{equation}
Therefore, 
the total heat current is
\begin{equation}
J= J_L =-J_R.
\end{equation}

The expression for heat conductivity is obtained considering the linear
response regime. Suppose that the $T_{L}=T+\Delta T/2$ and $T_{R}=T-\Delta T/2$, where 
$\Delta T$ is the temperature difference between left and right ends of the nanowire.
Performing Taylor expansion in $ \Delta T $
\begin{equation}
J=K \Delta T+....
\end{equation}
we identify the expression for the proportionality coefficient
\begin{equation}
K=\frac{1}{\hbar k_{B}T^{2}}\sum_{s}\epsilon_{s}^{2}\frac{\Gamma_{L}(\epsilon_{s})\Gamma_{R}(\epsilon_{s})}{\Gamma_{L}(\epsilon_{s})+\Gamma_{R}(\epsilon_{s})}\frac{e^{\epsilon_{s}/k_{B}T}}{(e^{\epsilon_{s}/k_{B}T}-1)^{2}}.
\end{equation}
Comparing  to Fourier's Law in one dimension 
\begin{equation}
\frac{J}{A}=\kappa \frac{\Delta T}{L},
\end{equation}
where $A$ is the cross-section area of the nanowire and $L$ is its length,
we can infer the expression for thermal conductivity is
\[
\kappa=\frac{KL}{A}.
\]

\section{Results}

\subsection{Model parameters}
\label{sec: model parameters}

For the results found throughout this paper the physically relevant quantities were calculated 
using molecular units (m.u.), which has energies of kJ/mol, for the numerical component of calculations \ref{numerical}.
Additionally for comparison to the experimental data for silicon nanowires there are a number of model parameters to consider. For the individual membranes the mechanical constant $c$ was taken to be the bulk counterpart, the speed of sound in silicon ($c=8433$ m/s). 

The  cut-off
for local vibrations $K_{\alpha}$ and the reservoir couplings
(\ref{eq:Reservoir Coupling}) were approximated by comparing the model to the experimental data from Li et. al. \cite{Li2003}. The comparison was made by calculating the conductivity-temperature profile for smooth nanowires and varying the cut-off until it provided a good fit across the different diameters, giving preference to better describing low temperatures (less than 150K). Additionally a reference energy of 1.1 kJ/mol is used to ensure that the phonon energies remain positive after diagonalization. The reservoir couplings were found by minimising the mean square residuals between the model and experiment.  The resulting couplings are roughly proportional to the cross-sectional area of the nanowires.
The choice of preferencing the low temperature fit is due to the lack of temperature dependent effects, particularly thermal expansion. As a result when the phonon transport is saturated the thermal conductivity plateaus whereas in the experimental results temperature dependent effects, such as thermal expansion, begin to dominate changes in the conductivity leading to a falloff in the in the conductivity at high temperatures. As a result of the decisions above we use the following set of fitted parameters in the majority of our calculations (for nanowire with diameter of 37 nm):
$K_\alpha=3.3$  kJ/mol and $\Gamma_L=\Gamma_R=23.1$ \, kJ/mol.

\subsection{Role of disorder (surface roughness)}
\begin{figure}[!ht]
	\begin{centering}
		(a)\includegraphics[scale=0.45]{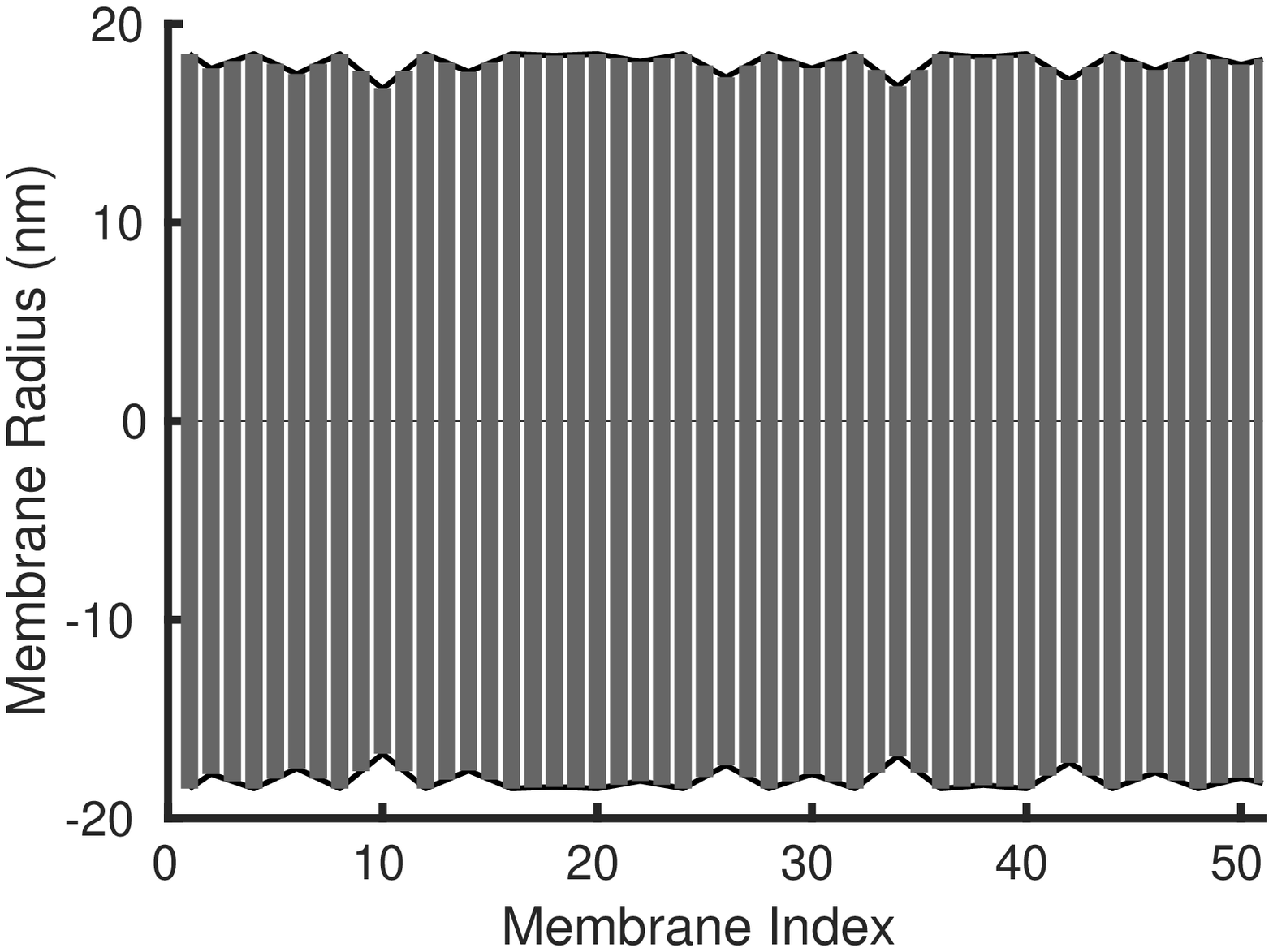} \label{figure: membrane lattice example a}
		(b)\includegraphics[scale=0.45]{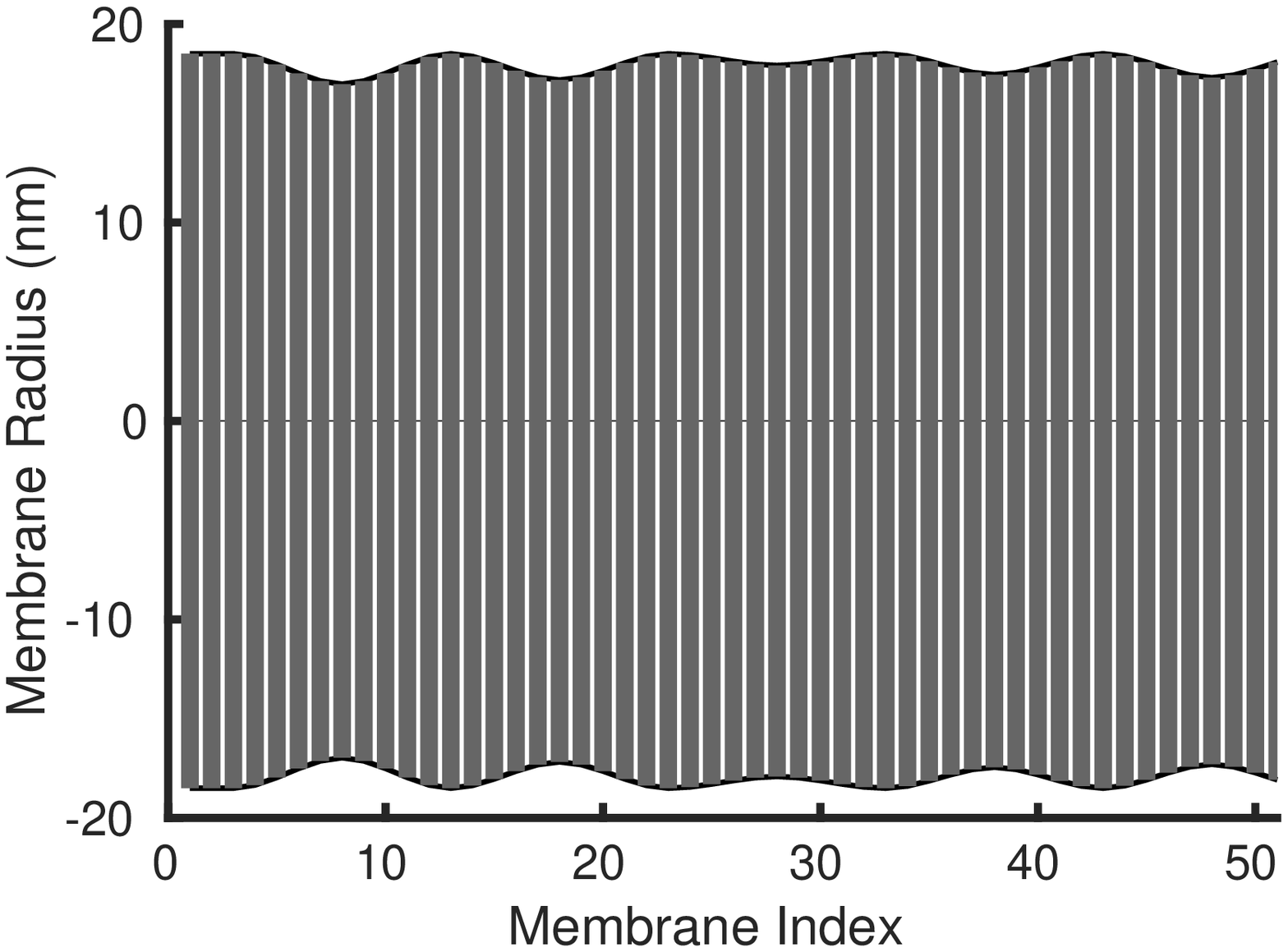} \label{figure: membrane lattice example b}
		\caption{Illustration of membranes making up two disordered nanowires a) with a corrugation length of $ \sigma_{L} = 4 $ and b) with a corrugation length of $\sigma_{L} = 10$. Both are 50 membranes long and have maximum diameter of $37$ nm and a roughness depth of $\sigma = 2$\ nm.}
		\label{figure: membrane lattice example}
	\end{centering}
\end{figure}

The surface roughness is introduced as a randomisation of the diameter of a subset of the membranes which make up the nanowire. This randomisation of the diameters change the local vibrational spectrum for each membrane (\ref{eq:Membrane H}). The radius of each of the membranes is randomized such that
\begin{equation}
R'_{\alpha}= R_{\alpha} - \Delta R_\alpha,
\end{equation}
where the $\Delta R_\alpha$ represents the magnitude of the disorder and $R_{\alpha}$ is the unmodified radius of the nanowire with $R'_\alpha$ representing the radius of disordered or "etched" nanowire. The disorder $\Delta R_\alpha$ is achieved by randomly sampling the uniform distribution between 0 and a roughness depth $\sigma$.  To introduce a tuneable corrugation length into the realisations, each of these membranes are placed half the desired corrugation length apart and the membranes that lie between them are interpolated to produce smoothed corrugations. Finally the membranes which are coupled directly to the driving reservoirs remain unmodified to ensure symmetry in the reservoir couplings \ref{eq:Reservoir Coupling}. Figure \ref{figure: membrane lattice example} shows examples of nanowires generated in this fashion. These realisations illustrate how the corrugation length affects the roughness profile.

\begin{figure}[!ht]
	\begin{centering}
		(a)\includegraphics[scale=0.45]{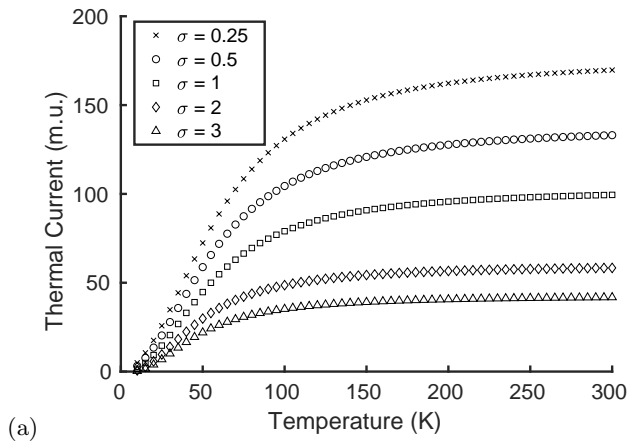}
		\caption{(a) Current-Temperature Profiles for various surface roughness depths $\sigma$ with corrugation length $ \sigma_{L} = 4 $ for a nanowire of diameter $D=37$ nm.}
		\label{figure: disorder depth effects}
	\end{centering}
\end{figure}

Throughout this paper for each set of system parameters observables are averaged over 500 realisations of disordered nanowires. The introduction of disorder results in a reduction of the thermal current of up two orders of magnitude with roughness less than half the membrane radius. This reduction is significant for even small amounts of disorder while further increases in disorder have less of an effect. Figure \ref{figure: disorder depth effects} illustrates this behaviour in the decreasing magnitude of the current with increasing disorder in the current-temperature profiles. The results clearly demonstrates the diminishing effect with increasing disorder depth by comparing the thermal current of nanowires at 300K for various roughness depths.

\begin{figure}[H]
	\begin{centering}
		(a)\includegraphics[scale=0.45]{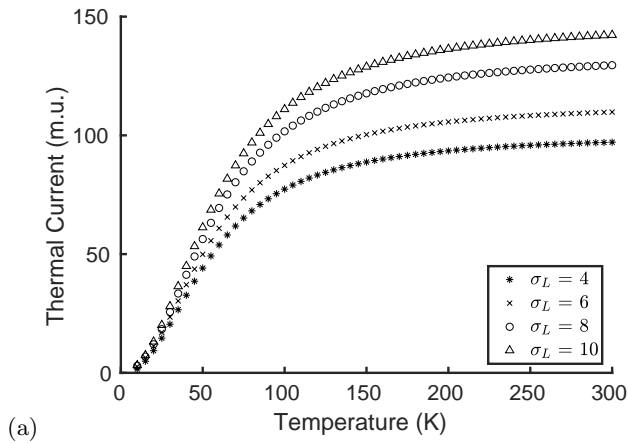}
		\caption{(a) Thermal current as a function of temperature for different roughness corrugation lengths. }
		\label{figure: disorder corrugation length effects}
	\end{centering}
\end{figure}

Conversely  as can be seen in figure \ref{figure: disorder corrugation length effects}, short corrugation lengths result in a significant reduction in thermal current, with thermal current increasing with corrugation length. This interplay between roughness depth and roughness corrugation length is qualitatively similar to what was observed in experimental results investigating the effects of correlation length on the thermal conductivity of vapour-liquid-solid (VLS) grown nanowires \cite{Lim2012}.

\subsubsection*{Contributions from individual normal modes}
As we discussed in the previous section, the nanowire surface roughness  results in an altered temperature profile where the thermal conductivity does not increase with temperature as much as in the smooth nanowires. This behaviour can be explained by looking at the contributions to the thermal current from the individual nanowire normal modes.
 In comparison to the smooth system, in the disordered systems higher energy phonons contribute less to thermal transport while lower energy levels are saturated leading to a shallower thermal current vs temperature profile. The shift towards lower energy phonons is due to disorder introducing a mismatching of local vibrational energies between neighbouring membranes. Disorder through the randomisation of the radii alters the dispersion of the local vibrational modes (\ref{eq:dispersion}). Differences in the dispersion of the local energy spectrum lead to smaller energy differences between the low energy vibrations of neighbouring membranes while at higher energies the mismatch is greater. This mismatch in neighbouring vibrational energies results in a weaker neighbouring coupling (\ref{eq:hopping}) and hence lowers transport.

\begin{figure}[H]
	\begin{centering}
		(a)\includegraphics[scale=0.4]{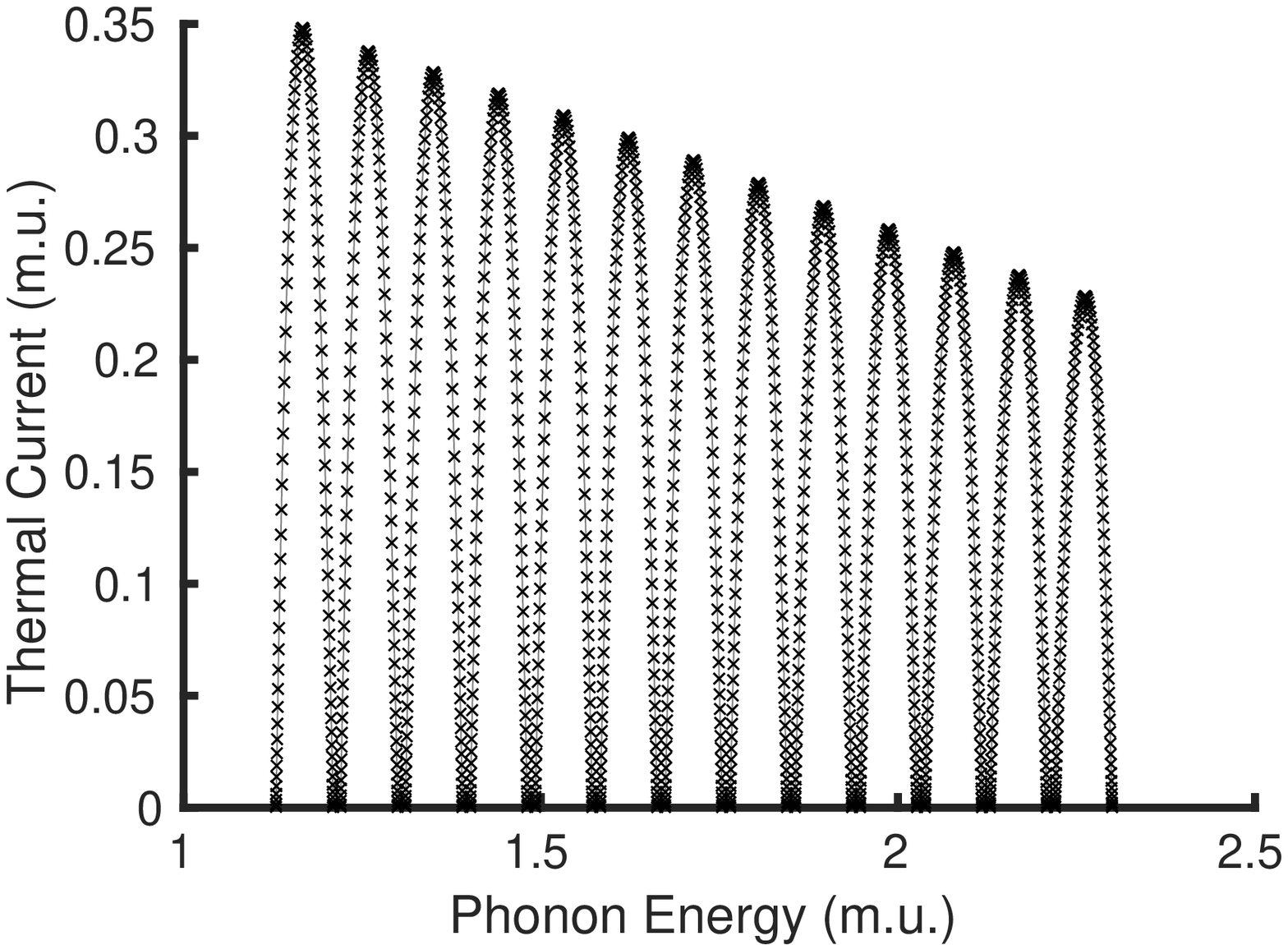}
		(b)\includegraphics[scale=0.4]{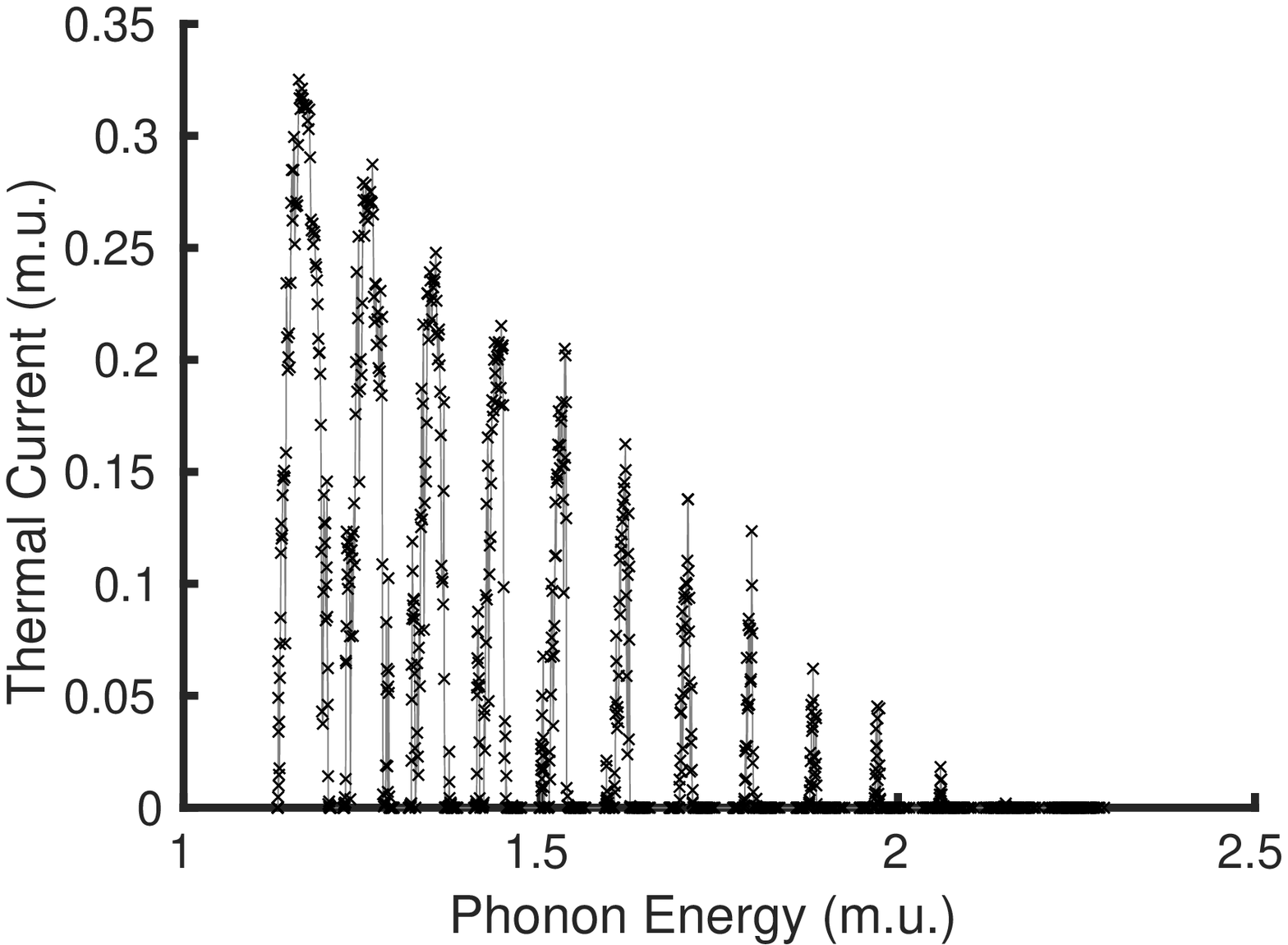}
		(c)\includegraphics[scale=0.4]{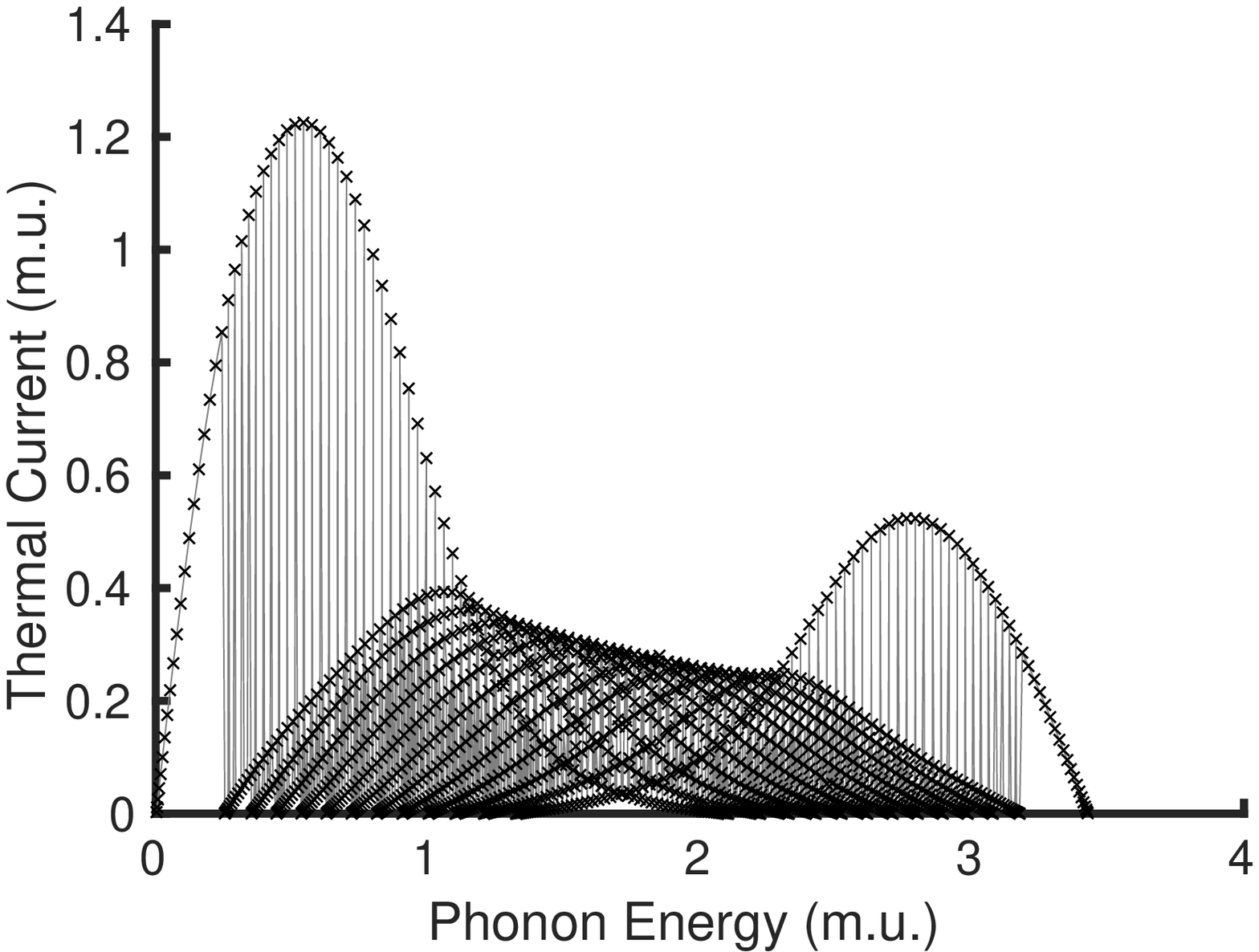}
		(d)\includegraphics[scale=0.4]{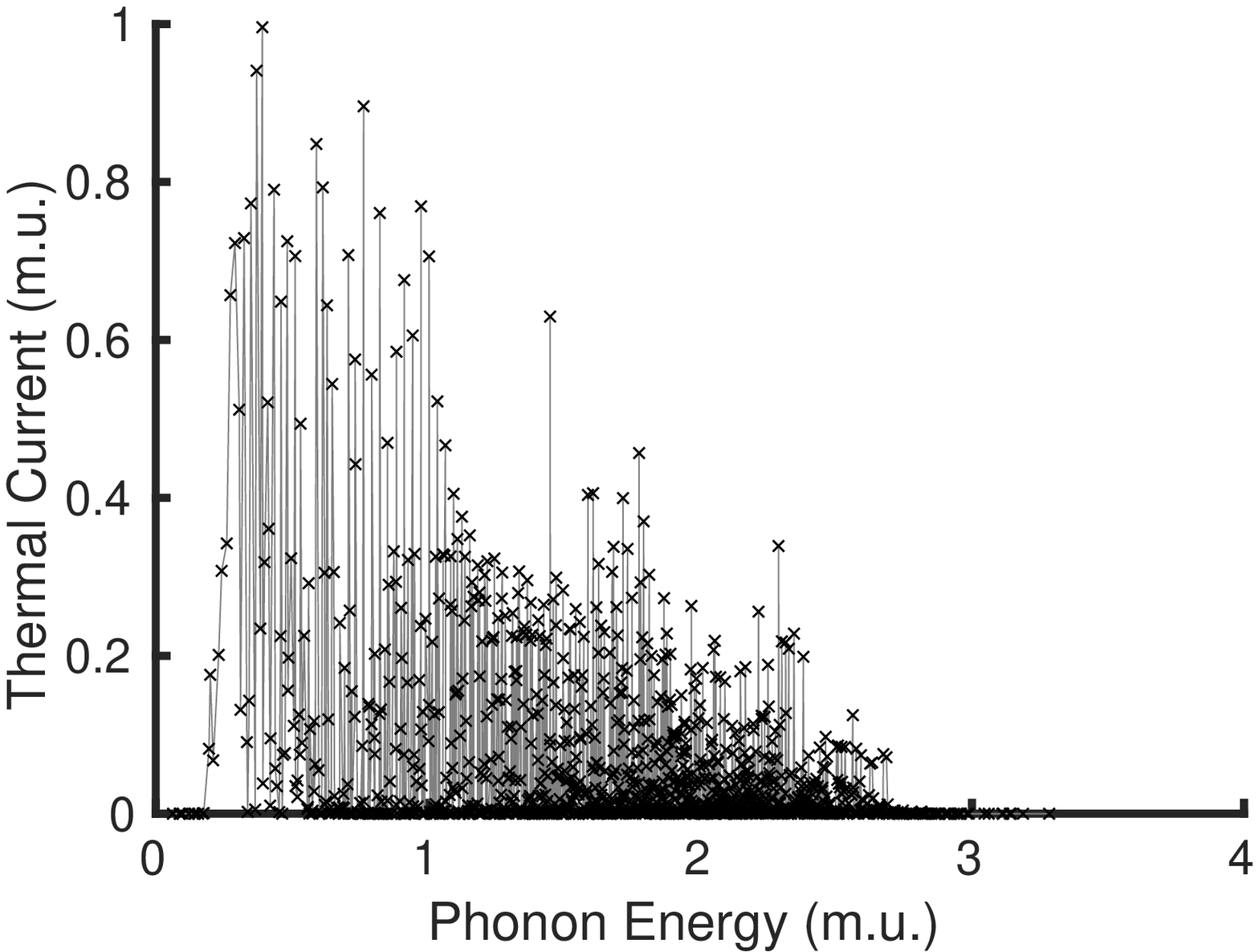}
		\caption{The contribution to the thermal flux from phonons of various energy for smooth nanowires a) and c); and disordered nanowires b) and d) $\sigma=0.75,\ \sigma_{L}=4$. Figures a) and b) have coupling parameter $v_0 = 0.01$ while c) and d) have coupling parameter $v_0 = 0.5$. Coupling parameter $ q = 0.01 $ for all sub-figures.}
		\label{figure: phonon subbands}
	\end{centering}
\end{figure}

Figure \ref{figure: phonon subbands} shows this effect for two different coupling regimes. The figure shows the contribution of various
phonon energies to the thermal current. For all four sub-figures \ref{figure: phonon subbands} (a)-(d) the nanowires have the same resonance parameter ($ q = 0.01 $) and the same "initial" radius ($ D = 37nm $) while sub-figures (b) and (d) have disordered radii. Sub-figures (a)  and (b) demonstrate that a weaker nearest neighbour coupling ($ v_0 = 0.01 $) results in separated and non-overlapping vibrational subbands in the nanowire vibrational spectra. These subbands are related to the one local vibrational modes contributed by each membrane. In this weak coupling regime local vibrational modes of one membrane only couple to local vibrational modes closest in energy. Thereby the lowest energy subband is the contribution of the lowest energy local vibrational mode from each of the membranes. When the coupling strength is increased local vibrational modes can couple to energy levels other than the closest energetically. Sub-figures (c) and (d) correspond to a stronger coupling ($ v_0=0.5 $). The coupling between more local vibrational modes leads to an overlapping of the subbands leaving no gaps.

Neither regime has an effect on the thermal current for smooth nanowires (without disorder). However the coupling parameters $ v_0 $ and $ q $ which govern the transition between the two regimes play an important role in determining how strong the effect of disorder on thermal transport in the system and consequently whether the system transports thermal energy in a more ballistic or wave like fashion or more diffusivity akin to Fourier's law.

\subsection{Length dependence of thermal current}

In a system with no disorder in the membrane diameter the system remains in a ballistic regime due to the  Hamiltonian (\ref{eq:NW Hamiltonian}) which does not  have phonon-phonon interactions in the longitudinal direction. The introduction of even a small amount of disorder into the membrane radii reduces the thermal conductivity by an order of magnitude and introduces a length dependence of the form $\kappa \propto L^{\beta}$ or similarly in the thermal current $ J \propto L^{\beta -1} $. This power law behaviour is indicative of disordered 1D lattices which conserve total phonon momentum \cite{Casher1971,Hu1998,Lepri1997,Hatano1999,lepri1998,Narayan2002} and in silicon nanowires using molecular dynamic simulations \cite{Yang2010} for systems below the phonon mean free path.  This is expected as the Hamiltonian (\ref{eq:NW Hamiltonian}) lacks an-harmonic phonon-phonon interaction terms and likewise conserves total phonon momentum. However unlike what has been reported previously for 1D momentum conserving systems the magnitude of $\beta$ is dependent on both the magnitude of the disorder and the corrugation length of the disorder.

In other 1D momentum conserving lattices the power law behaviour is dependent on the nature of the system interactions and reservoir couplings such as in disordered harmonic systems where a value of $\beta$ = 3/2 was found for free boundary conditions \cite{Dhar2001} and $\beta$ = 1/2 for fixed boundary conditions. This is independent of the magnitude or corrugation length of the disorder introduced. For anharmonic systems the power law behaviour is dependent on the nature of the anharmonicity. For example the diatomic Toda lattice has $\beta$ $\approx$ 0.35 \cite{Hatano1999} and the FPU chains exhibit $\beta$ = 1/3 when anharmonicity is strong. When disorder is more dominant than the anharmonicity the length dependence of the system is similar to the harmonic systems due to disorder suppressing the anharmonicity. In these other 1D systems the power law behaviour has a single characteristic parameter $ \beta $ with regards to disorder whereas in the thin membrane lattice model this behaviour is disorder dependant. From here the magnitude and corrugation length of the disorder as well as the inter-membrane coupling \ref{eq:hopping} are altered to demonstrate their effect on the power law dependence of the thermal current. The value of $ \beta $ is found using a least squares regression for lines of best fit on the numerical results. In addition to summarising this value, the tables also characterize the root mean squared roughness depth for the realisations.

\begin{figure}[H]
	\centering{}
	\includegraphics[scale=0.45]{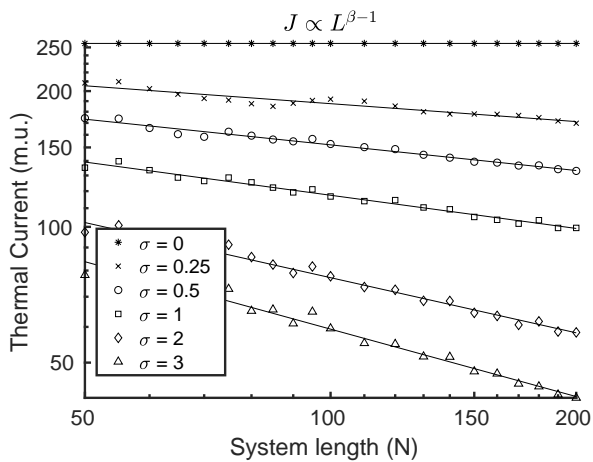}
	\caption{Log-log scale plot of the length dependence of thermal current
		at $T=300$K for various roughness depths. All curves have a
		corrugation length of $\sigma_{L}=1$nm and coupling parameters $v_0 = 0.5$ and $q = 0.01$.}
	\label{figure: length dependence - roughness depth}
\end{figure}

\begin{table}[H]
	\begin{centering}
		\begin{tabular}{|c|c|c|c|c|c|c|}
			\hline 
			Roughness & \multirow{2}{*}{0 nm } & \multirow{2}{*}{0.25 nm } & \multirow{2}{*}{0.5} & \multirow{2}{*}{1} & \multirow{2}{*}{2} & \multirow{2}{*}{3} \\
			Depth (nm) & & & & & & \\
			\hline
			\hline
			$\sqrt{\langle\sigma^2\rangle}$ & 0 & 0.087 & 0.17 & 0.35 & 0.70 & 1.04 \\
			\hline 
			$\beta$ & 1.000  & 0.868  & 0.811  & 0.755  & 0.594  & 0.503 \\
			\hline 
		\end{tabular}
		\par\end{centering}
	\caption{Values for $\beta$ corresponding to a power-law length
		dependence, $J= L^{\beta-1}$ of the thermal current for the
		various roughness depths in figure 6.
		$\sqrt{\langle\sigma^2\rangle}$ is the root mean squared roughness depth.}
	\label{table: length dependence - roughness depth}
\end{table}

Figure \ref{figure: length dependence - roughness depth} shows the length dependence of the thermal current for a series of $ D = 37 $ nm nanowires with different roughness depths but the same roughness corrugation length and coupling parameters. It illustrates how the power law behaviour is affected by the depth of the disorder. With increasing disorder leading to a more pronounced power law dependence of the thermal current. Table \ref{table: length dependence - roughness depth} summarizes the results in Figure \ref{figure: length dependence - roughness depth}, quantifying the value of $\beta$  for lines of best fit. In addition Table as well as the root mean square roughness depth for each set of realisations.

\subsubsection*{Roughness corrugation length}

The roughness corrugation length has a more significant impact on
the length dependence of the thermal conductivity than the roughness
depth.

\begin{figure}[H]
	\begin{centering}
		\includegraphics[scale=0.5]{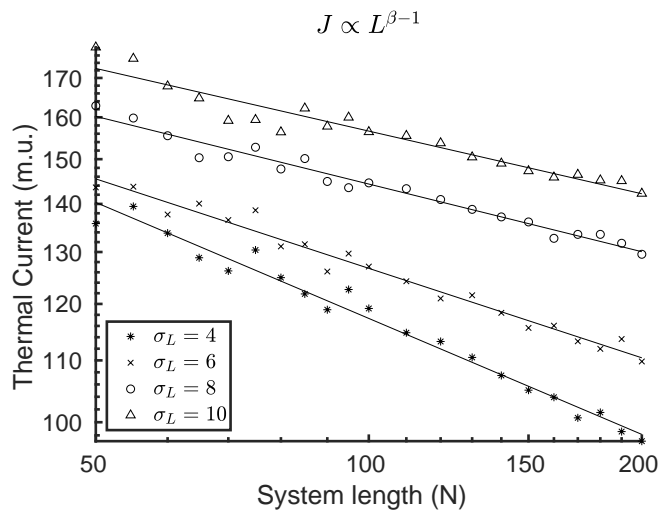} 
		\par\end{centering}
	\caption{Length dependence of thermal current at $T=300\,K$ for various
		roughness corrugation lengths. All curves have a roughness depth of
		$\sigma = 1\,nm$ and hopping parameters $v_0 = 0.5$ and $q = 0.01$.}
	\label{figure: length dependence - roughness corrugation length}
\end{figure}

\begin{table} [H]
	\begin{centering}
		\begin{tabular}{|c|c|c|c|c|}
			\hline 
			Roughness  & \multirow{2}{*}{4} & \multirow{2}{*}{6}  & \multirow{2}{*}{8}  & \multirow{2}{*}{10} \\
			corrugation Length &  &  &  & \\
			\hline
			\hline
			$\sqrt{\langle\sigma^2\rangle}$ & 0.35 & 0.35 & 0.34 & 0.34 \\
			\hline 
			$\beta$  & 0.742  & 0.801  & 0.850  & 0.860 \\
			\hline 
		\end{tabular}
		\par\end{centering}
	\caption{Values for $\beta$ corresponding to a power-law length
		dependence, $J = L^{\beta-1}$ of the thermal current for the
		various roughness corrugation lengths in Figure 7.}
	\label{table: length dependence - roughness corrugation length}
\end{table}

Figure \ref{figure: length dependence - roughness corrugation length} shows how increasing the corrugation length of the surface roughness increases the dependence of the thermal conductivity on length. All nanowires in the figure have an unaltered diameter of $ D = 37 $nm and have the same roughness depth and coupling parameters. Table \ref{table: length dependence - roughness corrugation length} summarizes this through quantification of $\beta$ as well as characterising the root mean squared roughness depth for each of the different corrugation length series.

\subsubsection*{Inter-membrane coupling}

The disorder induced length dependence is also influenced by the strength of the inter-membrane coupling (\ref{eq:hopping}) with a strong coupling increasing the length dependence. Lowering the parameter $q$ strengthens the coupling between non-equal local vibrational modes and increasing $v_0$ strengthens inter-membrane coupling generally. Each leading to a decrease in the value of $\beta.$

\begin{figure}[H]
	\begin{centering}
		\includegraphics[scale=0.5]{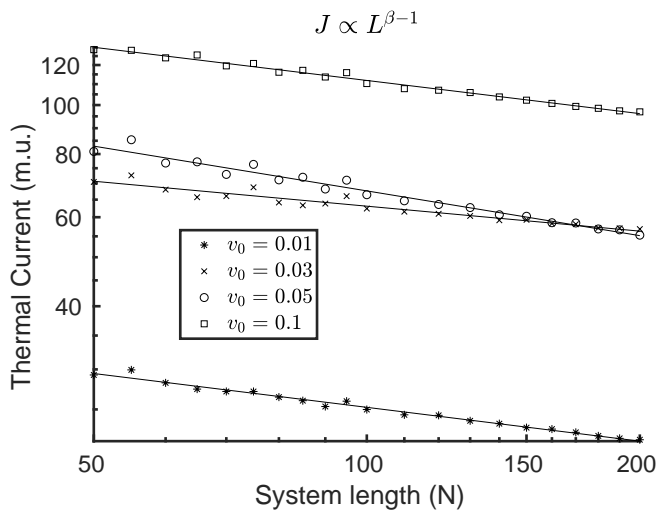} 
		\par\end{centering}
	\caption{Length dependence of thermal current at $T=300$K for various values for
		the coupling parameter $v_0$. All curves have a roughness depth of
		$\sigma=2$ nm and corrugation length of $\sigma_L = 4$ and
		coupling parameter $q = 0.01$.}
	\label{figure: length dependence - coupling para v0}
\end{figure}

\begin{table}[H]
	\begin{centering}
		\begin{tabular}{|c|c|c|c|c|}
			\hline 
			Coupling  & \multirow{2}{*}{0.01} & \multirow{2}{*}{0.03}  & \multirow{2}{*}{0.05}  & \multirow{2}{*}{0.1} \\
			Parameter $v_0$ &  &  &  & \\
			\hline
			\hline
			$\sqrt{\langle\sigma^2\rangle}$ & 0.69 & 0.69 & 0.70 & 0.69 \\
			\hline 
			$\beta$  & 0.778  & 0.836  & 0.706  & 0.782 \\
			\hline 
		\end{tabular}
		\par\end{centering}
	\caption{Values for $\beta$ corresponding to a power-law length
		dependence, $J = L^{\beta-1}$ of the thermal current for the
		various roughness corrugation lengths in Figure 8.}
	\label{table: length dependence - coupling para v0}
\end{table}

Figure \ref{figure: length dependence - coupling para v0} illustrates the length dependence of the thermal current on the coupling parameter $v_0$ for $ D = 37 $nm nanowires with the same surface roughness ($ \sigma = 2 $), roughness corrugation length ($ \sigma_{L} = 4 $) and resonance parameter ($ q = 0.01 $). Table III outlines the corresponding characteristics of the roughness and power law behaviour. An interesting feature in Figure \ref{figure: length dependence - coupling para v0} is the intersection of the $v_0$ = 0.03 and $v_0$ = 0.05 series indicating  the transition between the weak coupling regime and the strong coupling regime which corresponds to the transition from discrete bands in sub-figures \ref{figure: phonon subbands} (a) and (b) to the overlapping bands in sub-figures (c) and (d).

\begin{figure}[H]
	\begin{centering}
		\includegraphics[scale=0.5]{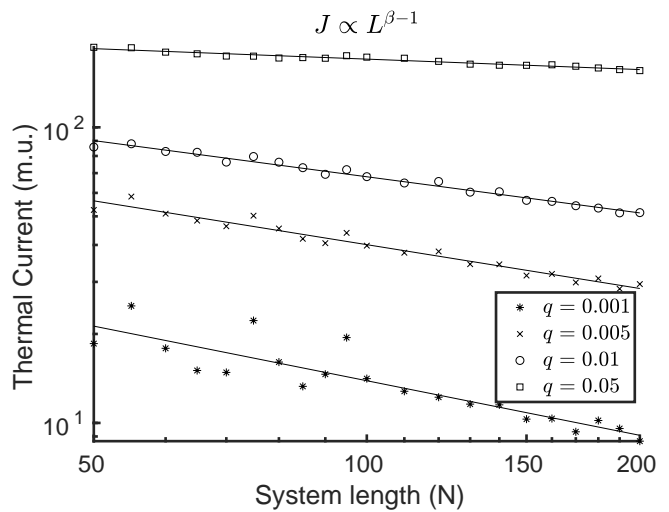} 
		\par\end{centering}
	\caption{Length dependence of thermal current at $T=300$K for various
		values of the coupling parameter $q$. All curves have a roughness depth of
		$\sigma=2$nm and a corrugation length $\sigma_L = 4$ and a coupling parameter
		$v_0 = 0.5$.}
	\label{figure: length dependence - coupling para q}
\end{figure}

\begin{table}[H]
	\begin{centering}
		\begin{tabular}{|c|c|c|c|c|}
			\hline 
			Coupling  & \multirow{2}{*}{0.001} & \multirow{2}{*}{0.005}  & \multirow{2}{*}{0.1}  & \multirow{2}{*}{0.5} \\
			Parameter $q$ &  &  &  & \\
			\hline
			\hline
			$\sqrt{\langle\sigma^2\rangle}$ & 0.70 & 0.70 & 0.69 & 0.69 \\
			\hline 
			$\beta$  & 0.386  & 0.508  & 0.594  & 0.882 \\
			\hline 
		\end{tabular}
		\par\end{centering}
	\caption{Values for $\alpha$ and $\beta$ corresponding to a power-law length
		dependence, $J = L^{\beta-1}$ of the thermal current for the
		various roughness corrugation lengths in Figure 9.}
	\label{table: length dependence - coupling para q}
\end{table}

Similarly Figure \ref{figure: length dependence - coupling para q} and Table \ref{table: length dependence - coupling para q} characterize the power law length dependence of the thermal conductivity and its relation to the coupling parameter $q$ for $ D = 37 $nm nanowires while keeping the same surface roughness ($ \sigma = 2 $), roughness corrugation length ($ \sigma_{L} = 4 $) and coupling strength ($ v_0 = 0.5 $). Value of the parameter q (\ref{eq:hopping}) indicates the relative  importance of resonance and off-resonance energy transfer. Increasing the parameter decreases the effect of energy mismatch between coupled energy levels and in turn decreases the effect that disorder has on the thermal current. Conversely decreasing the coupling parameter increases the energy mismatch and increases the effect of disorder.

Recent advances in the use of graphical processing units for computing have  allowed   molecular dynamics simulations of silicon nanowires to reach sizes comparable to the sizes in experiments \cite{Hou2016}. The simulations show that the thermal conductivity approaches a size independent regime ($ \beta = 0 $) as they increase in size. Our  membrane lattice model could suggest that the rate at which the nanowire approaches the size independent regime would depend on the nanowire material, through the inter-membrane coupling and the surface roughness characteristics.

\subsection{Comparison with  the experiment}

The thermal conductivity of nanowire  treated in the proposed model is reduced as the diameter of the nanowire is decreased similar to the experimental results for silicon nanowires \cite{Li2003}. However the temperature profile of the membrane lattice deviates from experimental profiles of silicon nanowires \cite{Li2003}. Several factors contribute to this discrepancy. The model relies on a simplified  local vibrational spectrum based on a classical elastic membrane with an introduced vibrational cutoff in a similar vein as a Debye cutoff. This local vibrational spectrum is also not temperature dependent, not accounting for thermal expansion.  In the experimental data this leads to a decrease in the thermal conductivity after the initial plateau which occurs between 100K and 150K which is not present in our thin-membrane lattice model.  This discrepancy is apparent when comparing to the original experiment by Li et al. \cite{Li2003} and less so when compared to the VLS grown nanowires in \cite{Lim2012} which is more subtle in the as grown nanowires.

\begin{figure}[!ht]
	\begin{centering}
		\includegraphics[bb=50bp 10bp 758bp 531bp,clip,scale=0.4]{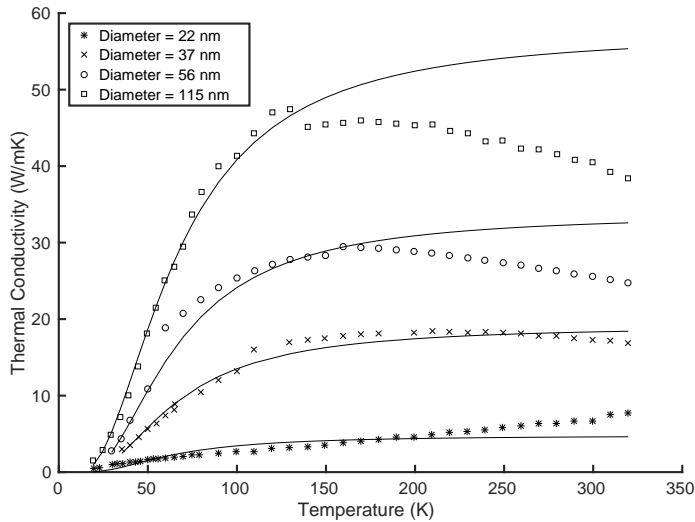}
		\par\end{centering}
	\caption{Shows thermal conductivity as a function of nanowire temperature (The
		average of the two driving reservoirs $\left(T_{L}+T_{R}\right)/2$).
		Diameters illustrated are 22nm, 37nm, 56nm, 115nm to draw comparison
		with the experimental data (dots) from Li et. al. 2003 \cite{Li2003}.}
	\label{figure: experimental fit}
\end{figure}

\begin{table}[H]
	\begin{centering}
		\begin{tabular}{|c|c|c|c|c|}
			\hline 
			Diameter $\left(nm\right)$ & 22 & 37 & 56 & 115\\
			\hline 
			Cut off $\left(k_{B}T\right)$ & 3.3 & 3.3 & 3.3 & 3.3\\
			\hline 
			Reservoir Coupling $\left(\gamma \right)$ & 3.8 & 23.1 & 64.0 & 218.1 \\
			\hline 
			$\gamma/A \times 10^{3}$ & 0.01 & 0.021 & 0.026 & 0.021\\
			\hline 
		\end{tabular}
		\par\end{centering}
	\caption{The values used to generate the fitted curves in figure 3. $ A $ represents the cross sectional area of the nanowire.}
	\label{table: fit parameters}
\end{table}

Figure \ref{figure: experimental fit} shows the fitted curves outlined in Section 3 A on model parameters and compares them to the experimental data from Li et al. \cite{Li2003} which they were fitted to. It also illustrates the discrepancy between the model and experiment at high temperatures. This  was due to the simplistic description for the local vibrational modes without taking into account the thermal expansion of the membrane radius. At these higher temperatures where phonon transport is saturated these effects dominate. This discrepancy is the motivating reason for preferencing the low temperature fit when developing the model parameters in section \ref{sec: model parameters}. Table \ref{table: fit parameters} presents a summary of the key parameters of the model which provides the fit to the experimental data and used to generate the model curves in Figure \ref{figure: experimental fit}. The table also illustrates that the reservoir couplings are proportional to the cross sectional areas of the nanowires.

\section{Conclusion}

We have developed a physical model for heat
transport in semiconductor nanowires. The nanowire is "sliced"  into
thin discs and the width of the disk is assumed to be much smaller than
radius of the nanowire, therefore the disc vibration spectrum
was approximated by the characteristic frequencies of vibrating thin elastic
membrane.
The left and right ends of the nanowire are attached to two macroscopic thermal baths held at different temperatures.
We treated the nanowire  as an open quantum system and derived  a Lindblad master equation for the nanowire density matrix.

Solving Linblad master equations for smooth and disordered nanowire we obtained key observables such as thermal conductivity coefficient and thermal flux.

	The main observations are as follows:
	\begin{itemize}		
		\item Thermal current is  significantly influenced not only by the nanowire surface disorder but also by the corrugation length of the disordered wire surface.
		The magnitude and corrugation length both play an important role in determining the magnitude and length dependence of the thermal current. An increase in the roughness depth from $ \sigma = 0.25$ nm to $ \sigma = 3$ nm resulted in a 100 m.u. reduction in the thermal current for nanowires at a temperature of $T=100$ K. Increasing the corrugation length of the surface roughness from four membranes ($\sigma_L = 4$) to ten ($\sigma_L = 10$) showed an increase in the thermal current from $ J=75 $ m.u. to $ J=110 $ m.u. for nanowires at $ T=100$ K.

		\item  The introduction of disorder does not stop the thermal conductivity from diverging  producing the power order length dependence of thermal conductivity $\kappa\propto L^\beta$. The value of $ \beta $ depends on the magnitude as well as corrugation length of the disorder. For, example, introducing a roughness depth of $ \sigma = 3$ nm	changes the ballistic transport regime, $ \beta = 1.0 $, to diffusive transport with significant length dependent thermal conductivity, $ \beta =  0.5$. For nanowires with a roughness depth the model demonstrated that doubling the corrugation length from four membranes to eight showed an increase in the length dependence parameter from $ \beta=0.74 $ to $ \beta = 0.85 $.
	
		\item The inter-membrane coupling also effects the length dependence of the thermal current. As the introduction of disorder moves the energy levels of neighbouring membranes away from resonance, the inter-membrane coupling determines to what extent off-resonance energy transfer contributes to the thermal current. Changing the value of the hopping parameter $v_0$ shows a small change in the length dependence of the system; between $ \beta = 0.7$ and $\beta = 0.84$ over a order of magnitude change in $v_0$. Whereas in comparison an order of magnitude change in the inter-membrane coupling parameter $q$ results in a change in the length dependence between $\beta = 0.88$ and $\beta = 0.51$.

		\item  The model is checked against the available experimental date. It is found that the model reproduces some of the qualitative effects of disorder that are present in semiconductor nanowires however the simplifying assumptions of the model limited its ability to reproduce the entire range of experimental results. On one hand the model qualitatively agrees with observations made previously about the interplay between disorder and the lateral length scale of the disorder. While on the other hand, the model does not account for thermal radius expansion and phonon-phonon interactions. This somehow limits its  practical use as a predictive model as well as its ability to investigate whether disorder effects the length dependence of the system for nanowires beyond the mean-free path of the phonons.

	\end{itemize}

\begin{acknowledgments}
We would like to thank Peter Stokes and Samuel Rudge for many valuable discussions.
\end{acknowledgments}

\section*{Author Contribution Statement}
JL and DK contributed equally throughout the process of developing the model, performing calculations, analysing results and writing the manuscript.

\newpage{}

\appendix

\section{Derivation of the master equation for quantum wire density matrix}

The open quantum systems is generically described by the following
Hamiltonian 
\begin{equation}
H=H_{S}+H_{B}+H_{SB},
\end{equation}
where $H_{S}$ is the system Hamiltonian, $H_{B}$ is the bath Hamiltonian,
and $H_{SB}$ is the system-bath interaction. The system-bath interaction
can be written as a product of operators in the system space $S_{i}$
and bath operators $B_{i}$: 
\begin{equation}
V=\sum_{\mu}S_{\mu}B_{\mu}=S_{\mu}B_{\mu}
\label{vsb-r}
\end{equation}

Using the Born-Markov approximation

\begin{align}
i\hbar\dot{\sigma}(t)= & [H_{S},\sigma(t)]\nonumber\\
& -\frac{i}{\hbar}\int_{0}^{\infty}d\tau\;\left[\left(G_{\mu\alpha}(\tau)S_{\mu}e^{-\frac{i}{\hbar}H_{S}\tau}S_{\alpha}e^{\frac{i}{\hbar}H_{S}\tau}\sigma(t)
\right. \right.\nonumber
\\
& \left. \left.
-G_{\alpha\mu}(-\tau)S_{\mu}\sigma(t)e^{-\frac{i}{\hbar}H_{S}\tau}S_{\alpha}e^{\frac{i}{\hbar}H_{S}\tau}\right)+\text{\text{h.c.}}\right]\label{master-eq} 
\end{align}

where the bath correlation function

\begin{equation}
G_{\mu\alpha}(\tau)=\text{Tr}_{B}\langle B_{\mu}e^{-i(H_{B}+H_{S})\tau/\hbar}B_{\alpha}\rangle.
\end{equation}

This master equation will be the starting point for our calculations.
This master equation (especially when it is written in the basis of
eigenstates of Hamiltonian $H_{S}$) is often called the \textit{Redfield master equation}. 
We represent the energy transfer  interaction between the nanowire and left and right thermal bath (\ref{vsb})
in the form suitable for the Redfield equation (\ref{vsb-r})
by introducing the set of 4 non-Hermitian operators

\begin{equation}
S_{1}=\sum_{s}v_{Ls}B_{s},\;\;\;S_{2}=\sum_{s}v_{Ls}B_{s}^{\dag},
\end{equation}

\begin{equation}
S_{3}=\sum_{s}v_{Rs}B_{s},\;\;\;S_{4}=\sum_{s}v_{Rs}B_{s}^{\dag},
\end{equation}

\begin{equation}
B_{1}=\sum_{l}b_{l}^{\dag},\;\;\;B_{2}=\sum_{l}b_{l},
\end{equation}

\begin{equation}
B_{3}=\sum_{r}b_{r}^{\dag},\;\;\;B_{4}=\sum_{r}b_{r}.
\end{equation}

The bath correlation function has the following nonzero
matrix elements $G_{12},G_{21},G_{34},G_{43}$, which can be easily
computed: correlation functions for the left bath are 

\begin{equation}
G_{12}(\tau)=\sum_{l}e^{\frac{i}{\hbar}\epsilon_{l}\tau}n_{l},
\end{equation}

\begin{equation}
G_{21}(\tau)=\sum_{l}e^{-\frac{i}{\hbar}\epsilon_{l}\tau}(1+n_{l}).
\end{equation}

and likewise, the correlation functions for the right bath are 

\begin{equation}
G_{34}(\tau)=\sum_{r}e^{\frac{i}{\hbar}\epsilon_{r}\tau}n_{r},
\end{equation}

\begin{equation}
G_{43}(\tau)=\sum_{r}e^{-\frac{i}{\hbar}\epsilon_{r}\tau}(1+n_{r}).
\end{equation}

Substituting these correlation functions and operators into the general Redfield master equation (\ref{master-eq}) and using rotating wave approximation we get

\begin{align}
i\hbar\dot{\rho}=&[(H_{W}+\sum_{s}\Delta_{L}(s)B_{s}^{\dagger}B_{s}+\sum_{s}\Delta_{R}(s)B_{s}^{\dagger}B_{s},\rho]\nonumber\\
&-i\sum_{s,\alpha=L,R}\Gamma_{\alpha}(s) \Big[\frac{1}{2}(1+2n_{\alpha}(s))\{B_{s}^{\dag}B_{s},\rho\}\nonumber\\
&\hskip 8em -(1+n_{\alpha}(s))B_{s}\rho B^{\dag}\nonumber\\
&\hskip 8em -n_{\alpha}(s)B_{s}^{\dag}\rho B_{s}+n_{\alpha}(s)\rho \Big].
\end{align}

Here the Lamb shifts due to left ($\alpha=L$) and right ($\alpha=R$) baths are

\begin{equation}
\Delta_{\alpha}(s)=[v_{\alpha s}]^{2}\sum_{k \in \alpha}\frac{\epsilon_{s}-\epsilon_{k}}{(\epsilon_{k}-\epsilon_{s})^{2}+\nu^{2}}
\end{equation}

and the level broadening functions are

\begin{equation}
\Gamma_{\alpha}(s)=2\pi[v_{\alpha s}]^{2}\rho_{\alpha}(s).
\end{equation}

\section{Practical calculations}
 Outlined here are the key steps for practical calculations to obtain observables from the model.

Inputs: $T$ - temperature , $N$ - length (number of membranes),
$R_{i}\;\;\;i=1,...,N$ radius, $v_0, q$ - coupling between membranes,
$\gamma_{L/R}=2\pi t^{2}\rho_{L/R}$ - couplings between the end membranes and reservoirs.
\begin{enumerate}
	\item Set up the Hamiltonian for the wire. Compute frequencies of intrinsic
	vibrations of the membrane ($\alpha=1,..,N$, $k=1,....,K_{\alpha}$)
	\begin{equation}
	E_{\alpha k}=\hbar\lambda_{\alpha k}c,
	\end{equation}
	where $\lambda_{\alpha k}$ is determined from zeroes of Bessel function
	$J_{0}(\lambda_{\alpha k}R_{\alpha})=0$. Here $K_{\alpha}$ is the
	natural cut-off for internal vibrations of membrane $\alpha$. This
	natural cutoff is in analogue to a Debye cutoff introducing the discrete
	atomic structure and its restriction on higher energy modes. 
	\begin{align}
	H_{W}=&\sum_{\alpha}^{N}\sum_{k=1}^{K_{\alpha}}E_{\alpha k}b_{\alpha k}^{\dagger}b_{\alpha k}\nonumber\\
	&+v\sum_{\alpha}^{N-1}\sum_{kk'}\left[b_{\alpha k}^{\dagger}b_{\alpha+1k'}+b_{\alpha+1k}^{\dagger}b_{\alpha k'}\right].
	\end{align}
	\item Form and diagonalize the Hamiltonian matrix
	Form matrix $h_{\alpha k,\alpha'k'}$ --- it has dimension $(K_{1}K_{2}...K_{N})\times(K_{1}K_{2}...K_{N})$
	\begin{equation}
	H_{W}=\sum_{\alpha k}\sum_{\alpha'k'}h_{\alpha k,\alpha'k'}b_{\alpha k}^{\dagger}b_{\alpha'k'}\label{eq:Quadratic H}
	\end{equation}
	Diagonalize the Hamiltonian by solving the eigenvalue problem 
	\begin{equation}
	\sum_{\alpha'k'}h_{\alpha k,\alpha'k'}U_{\ \alpha'k',s}=\epsilon_{s}U_{\alpha k,s}, \label{numerical}
	\end{equation}
	\item Evaluate the master equation normalisation and dissipation terms in
	the broad band limit so that $\Delta_{L,R}(s)\rightarrow0$ and the
	reservoir couplings are 
	\begin{equation}
	\Gamma_{L}(s)=\gamma_{L}\sum_{k=1}^{K_{1}}U_{1k,s}\;\;\;\;\Gamma_{R}(s)=\gamma_{R}\sum_{k=1}^{K_{N}}U_{Nk,s}\label{eq:Reservoir Coupling}
	\end{equation}
	\item Compute heat conductivity and other observables using analytical formulae
	($n_{S}=K_{1}K_{2}....K_{N}$) 
	\begin{equation}
	\kappa=\frac{L}{\hbar k_{B}T^{2} A}\sum_{s=1}^{n_{S}}\epsilon_{s}^{2}\frac{\Gamma_{L}(s)\Gamma_{R}(s)}{\Gamma_{L}(s)+\Gamma_{R}(s)}\frac{e^{\epsilon_{s}/k_{B}T}}{(e^{\epsilon_{s}/k_{B}T}-1)^{2}}
	\end{equation}
\end{enumerate}
\newpage
\bibliographystyle{apsrev4-1}
\phantomsection\addcontentsline{toc}{section}{\refname}\bibliography{References}

\begin{thebibliography}{28}%
\makeatletter
\providecommand \@ifxundefined [1]{%
 \@ifx{#1\undefined}
}%
\providecommand \@ifnum [1]{%
 \ifnum #1\expandafter \@firstoftwo
 \else \expandafter \@secondoftwo
 \fi
}%
\providecommand \@ifx [1]{%
 \ifx #1\expandafter \@firstoftwo
 \else \expandafter \@secondoftwo
 \fi
}%
\providecommand \natexlab [1]{#1}%
\providecommand \enquote  [1]{``#1''}%
\providecommand \bibnamefont  [1]{#1}%
\providecommand \bibfnamefont [1]{#1}%
\providecommand \citenamefont [1]{#1}%
\providecommand \href@noop [0]{\@secondoftwo}%
\providecommand \href [0]{\begingroup \@sanitize@url \@href}%
\providecommand \@href[1]{\@@startlink{#1}\@@href}%
\providecommand \@@href[1]{\endgroup#1\@@endlink}%
\providecommand \@sanitize@url [0]{\catcode `\\12\catcode `\$12\catcode
  `\&12\catcode `\#12\catcode `\^12\catcode `\_12\catcode `\%12\relax}%
\providecommand \@@startlink[1]{}%
\providecommand \@@endlink[0]{}%
\providecommand \url  [0]{\begingroup\@sanitize@url \@url }%
\providecommand \@url [1]{\endgroup\@href {#1}{\urlprefix }}%
\providecommand \urlprefix  [0]{URL }%
\providecommand \Eprint [0]{\href }%
\providecommand \doibase [0]{http://dx.doi.org/}%
\providecommand \selectlanguage [0]{\@gobble}%
\providecommand \bibinfo  [0]{\@secondoftwo}%
\providecommand \bibfield  [0]{\@secondoftwo}%
\providecommand \translation [1]{[#1]}%
\providecommand \BibitemOpen [0]{}%
\providecommand \bibitemStop [0]{}%
\providecommand \bibitemNoStop [0]{.\EOS\space}%
\providecommand \EOS [0]{\spacefactor3000\relax}%
\providecommand \BibitemShut  [1]{\csname bibitem#1\endcsname}%
\let\auto@bib@innerbib\@empty
\bibitem [{\citenamefont {Rieder}\ \emph {et~al.}(1967)\citenamefont {Rieder},
  \citenamefont {Lebowitz},\ and\ \citenamefont {Lieb}}]{Rieder1967}%
  \BibitemOpen
  \bibfield  {author} {\bibinfo {author} {\bibfnamefont {Z.}~\bibnamefont
  {Rieder}}, \bibinfo {author} {\bibfnamefont {J.~L.}\ \bibnamefont
  {Lebowitz}}, \ and\ \bibinfo {author} {\bibfnamefont {E.}~\bibnamefont
  {Lieb}},\ }\href {\doibase 10.1063/1.1705319} {\bibfield  {journal} {\bibinfo
   {journal} {Journal of Mathematical Physics}\ }\textbf {\bibinfo {volume}
  {8}},\ \bibinfo {pages} {1073} (\bibinfo {year} {1967})}\BibitemShut
  {NoStop}%
\bibitem [{\citenamefont {Peshkov}(1944)}]{Peshkov1944}%
  \BibitemOpen
  \bibfield  {author} {\bibinfo {author} {\bibfnamefont {V.}~\bibnamefont
  {Peshkov}},\ }\href@noop {} {\bibfield  {journal} {\bibinfo  {journal} {J.
  Phys. (Moscow)}\ }\textbf {\bibinfo {volume} {8}} (\bibinfo {year}
  {1944})}\BibitemShut {NoStop}%
\bibitem [{\citenamefont {Guyer}\ and\ \citenamefont
  {Krumhansl}(1966)}]{Guyer1966}%
  \BibitemOpen
  \bibfield  {author} {\bibinfo {author} {\bibfnamefont {R.~A.}\ \bibnamefont
  {Guyer}}\ and\ \bibinfo {author} {\bibfnamefont {J.~A.}\ \bibnamefont
  {Krumhansl}},\ }\href {\doibase 10.1103/PhysRev.148.778} {\bibfield
  {journal} {\bibinfo  {journal} {Physical Review}\ }\textbf {\bibinfo {volume}
  {148}},\ \bibinfo {pages} {778} (\bibinfo {year} {1966})}\BibitemShut
  {NoStop}%
\bibitem [{\citenamefont {Casher}\ and\ \citenamefont {J}(1971)}]{Casher1971}%
  \BibitemOpen
  \bibfield  {author} {\bibinfo {author} {\bibfnamefont {A.}~\bibnamefont
  {Casher}}\ and\ \bibinfo {author} {\bibfnamefont {L.}~\bibnamefont {J}},\
  }\href {\doibase 10.1063/1.1665794} {\bibfield  {journal} {\bibinfo
  {journal} {Journal of Mathematical Physics}\ }\textbf {\bibinfo {volume}
  {12}},\ \bibinfo {pages} {1701} (\bibinfo {year} {1971})}\BibitemShut
  {NoStop}%
\bibitem [{\citenamefont {Hu}\ \emph {et~al.}(1998)\citenamefont {Hu},
  \citenamefont {Li},\ and\ \citenamefont {Zhao}}]{Hu1998}%
  \BibitemOpen
  \bibfield  {author} {\bibinfo {author} {\bibfnamefont {B.}~\bibnamefont
  {Hu}}, \bibinfo {author} {\bibfnamefont {B.}~\bibnamefont {Li}}, \ and\
  \bibinfo {author} {\bibfnamefont {H.}~\bibnamefont {Zhao}},\ }\href {\doibase
  10.1103/PhysRevE.57.2992} {\bibfield  {journal} {\bibinfo  {journal}
  {Physical Review E: Statistical, Nonlinear, and Soft Matter Physics}\
  }\textbf {\bibinfo {volume} {57}},\ \bibinfo {pages} {2992} (\bibinfo {year}
  {1998})}\BibitemShut {NoStop}%
\bibitem [{\citenamefont {Lepri}\ \emph {et~al.}(1997)\citenamefont {Lepri},
  \citenamefont {Livi},\ and\ \citenamefont {Politi}}]{Lepri1997}%
  \BibitemOpen
  \bibfield  {author} {\bibinfo {author} {\bibfnamefont {S.}~\bibnamefont
  {Lepri}}, \bibinfo {author} {\bibfnamefont {R.}~\bibnamefont {Livi}}, \ and\
  \bibinfo {author} {\bibfnamefont {A.}~\bibnamefont {Politi}},\ }\href
  {\doibase 10.1103/PhysRevLett.78.1896} {\bibfield  {journal} {\bibinfo
  {journal} {Physical Review Letters}\ }\textbf {\bibinfo {volume} {78}},\
  \bibinfo {pages} {1896} (\bibinfo {year} {1997})}\BibitemShut {NoStop}%
\bibitem [{\citenamefont {Hatano}(1999)}]{Hatano1999}%
  \BibitemOpen
  \bibfield  {author} {\bibinfo {author} {\bibfnamefont {T.}~\bibnamefont
  {Hatano}},\ }\href {\doibase 10.1103/PhysRevE.59.R1} {\bibfield  {journal}
  {\bibinfo  {journal} {Physical Review E: Statistical, Nonlinear, and Soft
  Matter Physics}\ }\textbf {\bibinfo {volume} {59}},\ \bibinfo {pages} {1}
  (\bibinfo {year} {1999})}\BibitemShut {NoStop}%
\bibitem [{\citenamefont {Lepri}\ \emph {et~al.}(1998)\citenamefont {Lepri},
  \citenamefont {Livi},\ and\ \citenamefont {Politi}}]{lepri1998}%
  \BibitemOpen
  \bibfield  {author} {\bibinfo {author} {\bibfnamefont {S.}~\bibnamefont
  {Lepri}}, \bibinfo {author} {\bibfnamefont {R.}~\bibnamefont {Livi}}, \ and\
  \bibinfo {author} {\bibfnamefont {A.}~\bibnamefont {Politi}},\ }\href
  {\doibase 10.1016/S0167-2789(98)00076-1} {\bibfield  {journal} {\bibinfo
  {journal} {Physica D: Nonlinear Phenomena}\ }\textbf {\bibinfo {volume}
  {119}},\ \bibinfo {pages} {140} (\bibinfo {year} {1998})}\BibitemShut
  {NoStop}%
\bibitem [{\citenamefont {Narayan}\ and\ \citenamefont
  {Ramaswamy}(2002)}]{Narayan2002}%
  \BibitemOpen
  \bibfield  {author} {\bibinfo {author} {\bibfnamefont {O.}~\bibnamefont
  {Narayan}}\ and\ \bibinfo {author} {\bibfnamefont {S.}~\bibnamefont
  {Ramaswamy}},\ }\href {\doibase 10.1103/physrevlett.89.200601} {\bibfield
  {journal} {\bibinfo  {journal} {Physical Review Letters}\ }\textbf {\bibinfo
  {volume} {89}},\ \bibinfo {pages} {200601} (\bibinfo {year}
  {2002})}\BibitemShut {NoStop}%
\bibitem [{\citenamefont {Li}\ \emph {et~al.}(2003)\citenamefont {Li},
  \citenamefont {Wu}, \citenamefont {Kim}, \citenamefont {Shi}, \citenamefont
  {Yang},\ and\ \citenamefont {Majumdar}}]{Li2003}%
  \BibitemOpen
  \bibfield  {author} {\bibinfo {author} {\bibfnamefont {D.}~\bibnamefont
  {Li}}, \bibinfo {author} {\bibfnamefont {Y.}~\bibnamefont {Wu}}, \bibinfo
  {author} {\bibfnamefont {P.}~\bibnamefont {Kim}}, \bibinfo {author}
  {\bibfnamefont {L.}~\bibnamefont {Shi}}, \bibinfo {author} {\bibfnamefont
  {P.}~\bibnamefont {Yang}}, \ and\ \bibinfo {author} {\bibfnamefont
  {A.}~\bibnamefont {Majumdar}},\ }\href {\doibase 10.1063/1.1616981}
  {\bibfield  {journal} {\bibinfo  {journal} {Applied Physics Letters}\
  }\textbf {\bibinfo {volume} {83}},\ \bibinfo {pages} {2934} (\bibinfo {year}
  {2003})}\BibitemShut {NoStop}%
\bibitem [{\citenamefont {Humphrey}\ and\ \citenamefont
  {Linke}(2005)}]{Humphrey2005}%
  \BibitemOpen
  \bibfield  {author} {\bibinfo {author} {\bibfnamefont {T.~E.}\ \bibnamefont
  {Humphrey}}\ and\ \bibinfo {author} {\bibfnamefont {H.}~\bibnamefont
  {Linke}},\ }\href {\doibase 10.1103/PhysRevLett.94.096601} {\bibfield
  {journal} {\bibinfo  {journal} {Physical Review Letters}\ }\textbf {\bibinfo
  {volume} {94}},\ \bibinfo {pages} {096601} (\bibinfo {year}
  {2005})}\BibitemShut {NoStop}%
\bibitem [{\citenamefont {Hochbaum}\ \emph {et~al.}(2008)\citenamefont
  {Hochbaum}, \citenamefont {Chen}, \citenamefont {Delago}, \citenamefont
  {Liang}, \citenamefont {Garnett}, \citenamefont {Najarian}, \citenamefont
  {Majumdar},\ and\ \citenamefont {Yang}}]{Hochbaum2008}%
  \BibitemOpen
  \bibfield  {author} {\bibinfo {author} {\bibfnamefont {A.}~\bibnamefont
  {Hochbaum}}, \bibinfo {author} {\bibfnamefont {R.}~\bibnamefont {Chen}},
  \bibinfo {author} {\bibfnamefont {R.~D.}\ \bibnamefont {Delago}}, \bibinfo
  {author} {\bibfnamefont {W.}~\bibnamefont {Liang}}, \bibinfo {author}
  {\bibfnamefont {E.~C.}\ \bibnamefont {Garnett}}, \bibinfo {author}
  {\bibfnamefont {M.}~\bibnamefont {Najarian}}, \bibinfo {author}
  {\bibfnamefont {A.}~\bibnamefont {Majumdar}}, \ and\ \bibinfo {author}
  {\bibfnamefont {P.}~\bibnamefont {Yang}},\ }\href {\doibase
  10.1038/nature06381} {\bibfield  {journal} {\bibinfo  {journal} {Nature}\
  }\textbf {\bibinfo {volume} {451}},\ \bibinfo {pages} {163} (\bibinfo {year}
  {2008})}\BibitemShut {NoStop}%
\bibitem [{\citenamefont {Mingo}\ and\ \citenamefont {Yang}(2003)}]{Mingo2003}%
  \BibitemOpen
  \bibfield  {author} {\bibinfo {author} {\bibfnamefont {N.}~\bibnamefont
  {Mingo}}\ and\ \bibinfo {author} {\bibfnamefont {L.}~\bibnamefont {Yang}},\
  }\href {\doibase 10.1021/nl034721i} {\bibfield  {journal} {\bibinfo
  {journal} {Nano Letters}\ }\textbf {\bibinfo {volume} {3}},\ \bibinfo {pages}
  {1713} (\bibinfo {year} {2003})}\BibitemShut {NoStop}%
\bibitem [{\citenamefont {Liang}\ and\ \citenamefont {Li}(2006)}]{Liang2006}%
  \BibitemOpen
  \bibfield  {author} {\bibinfo {author} {\bibfnamefont {L.}~\bibnamefont
  {Liang}}\ and\ \bibinfo {author} {\bibfnamefont {B.}~\bibnamefont {Li}},\
  }\href {\doibase 10.1103/PhysRevB.73.153303} {\bibfield  {journal} {\bibinfo
  {journal} {Physical Review B}\ }\textbf {\bibinfo {volume} {73}},\ \bibinfo
  {pages} {153303} (\bibinfo {year} {2006})}\BibitemShut {NoStop}%
\bibitem [{\citenamefont {Moore}\ \emph {et~al.}(2008)\citenamefont {Moore},
  \citenamefont {Saha}, \citenamefont {Prasher},\ and\ \citenamefont
  {Shi}}]{Moore2008}%
  \BibitemOpen
  \bibfield  {author} {\bibinfo {author} {\bibfnamefont {A.~L.}\ \bibnamefont
  {Moore}}, \bibinfo {author} {\bibfnamefont {S.~K.}\ \bibnamefont {Saha}},
  \bibinfo {author} {\bibfnamefont {R.~S.}\ \bibnamefont {Prasher}}, \ and\
  \bibinfo {author} {\bibfnamefont {L.}~\bibnamefont {Shi}},\ }\href {\doibase
  10.1063/1.2970044} {\bibfield  {journal} {\bibinfo  {journal} {Applied
  Physics Letters}\ }\textbf {\bibinfo {volume} {93}},\ \bibinfo {pages}
  {083112} (\bibinfo {year} {2008})}\BibitemShut {NoStop}%
\bibitem [{\citenamefont {Donadio}\ and\ \citenamefont
  {Galli}(2009)}]{Donadio2009}%
  \BibitemOpen
  \bibfield  {author} {\bibinfo {author} {\bibfnamefont {D.}~\bibnamefont
  {Donadio}}\ and\ \bibinfo {author} {\bibfnamefont {G.}~\bibnamefont
  {Galli}},\ }\href {\doibase 10.1103/PhysRevLett.102.195901} {\bibfield
  {journal} {\bibinfo  {journal} {Physical Review Letters}\ }\textbf {\bibinfo
  {volume} {102}},\ \bibinfo {pages} {195901} (\bibinfo {year}
  {2009})}\BibitemShut {NoStop}%
\bibitem [{\citenamefont {Martin}\ \emph {et~al.}(2009)\citenamefont {Martin},
  \citenamefont {Aksamija}, \citenamefont {Pop},\ and\ \citenamefont
  {Rabaioli}}]{Martin2009}%
  \BibitemOpen
  \bibfield  {author} {\bibinfo {author} {\bibfnamefont {P.}~\bibnamefont
  {Martin}}, \bibinfo {author} {\bibfnamefont {Z.}~\bibnamefont {Aksamija}},
  \bibinfo {author} {\bibfnamefont {E.}~\bibnamefont {Pop}}, \ and\ \bibinfo
  {author} {\bibfnamefont {U.}~\bibnamefont {Rabaioli}},\ }\href {\doibase
  10.1103/PhysRevLett.102.125503} {\bibfield  {journal} {\bibinfo  {journal}
  {Physical Review Letters}\ }\textbf {\bibinfo {volume} {102}},\ \bibinfo
  {pages} {125503} (\bibinfo {year} {2009})}\BibitemShut {NoStop}%
\bibitem [{\citenamefont {Kosevich}\ and\ \citenamefont
  {Savin}(2009)}]{Kosevich2009}%
  \BibitemOpen
  \bibfield  {author} {\bibinfo {author} {\bibfnamefont {Y.~A.}\ \bibnamefont
  {Kosevich}}\ and\ \bibinfo {author} {\bibfnamefont {A.~V.}\ \bibnamefont
  {Savin}},\ }\href {http://stacks.iop.org/0295-5075/88/i=1/a=14002} {\bibfield
   {journal} {\bibinfo  {journal} {EPL (Europhysics Letters)}\ }\textbf
  {\bibinfo {volume} {88}},\ \bibinfo {pages} {14002} (\bibinfo {year}
  {2009})}\BibitemShut {NoStop}%
\bibitem [{\citenamefont {Luisier}(2011)}]{Luisier2011}%
  \BibitemOpen
  \bibfield  {author} {\bibinfo {author} {\bibfnamefont {M.}~\bibnamefont
  {Luisier}},\ }\href {\doibase 10.1063/1.3644993} {\bibfield  {journal}
  {\bibinfo  {journal} {Journal of Applied Physics}\ }\textbf {\bibinfo
  {volume} {110}},\ \bibinfo {pages} {074510} (\bibinfo {year}
  {2011})}\BibitemShut {NoStop}%
\bibitem [{\citenamefont {Blanc}\ \emph {et~al.}(2013)\citenamefont {Blanc},
  \citenamefont {Rajabpour}, \citenamefont {Volz}, \citenamefont {Fournier},\
  and\ \citenamefont {Bourgeois}}]{Blanc2013}%
  \BibitemOpen
  \bibfield  {author} {\bibinfo {author} {\bibfnamefont {C.}~\bibnamefont
  {Blanc}}, \bibinfo {author} {\bibfnamefont {A.}~\bibnamefont {Rajabpour}},
  \bibinfo {author} {\bibfnamefont {S.}~\bibnamefont {Volz}}, \bibinfo {author}
  {\bibfnamefont {T.}~\bibnamefont {Fournier}}, \ and\ \bibinfo {author}
  {\bibfnamefont {O.}~\bibnamefont {Bourgeois}},\ }\href {\doibase
  10.1063/1.4816590} {\bibfield  {journal} {\bibinfo  {journal} {Applied
  Physics Letters}\ }\textbf {\bibinfo {volume} {103}},\ \bibinfo {pages}
  {043109} (\bibinfo {year} {2013})}\BibitemShut {NoStop}%
\bibitem [{\citenamefont {Carrete}\ \emph {et~al.}(2011)\citenamefont
  {Carrete}, \citenamefont {Gallego},\ and\ \citenamefont
  {Varela}}]{Carrete2011}%
  \BibitemOpen
  \bibfield  {author} {\bibinfo {author} {\bibfnamefont {J.}~\bibnamefont
  {Carrete}}, \bibinfo {author} {\bibfnamefont {L.~J.}\ \bibnamefont
  {Gallego}}, \ and\ \bibinfo {author} {\bibfnamefont {L.~M.}\ \bibnamefont
  {Varela}},\ }\href {\doibase 10.1103/PhysRevB.84.075403} {\bibfield
  {journal} {\bibinfo  {journal} {Physical Review B: Condensed Matter and
  Materials Physics}\ }\textbf {\bibinfo {volume} {84}},\ \bibinfo {pages}
  {075403} (\bibinfo {year} {2011})}\BibitemShut {NoStop}%
\bibitem [{\citenamefont {Sadhu}\ and\ \citenamefont
  {Sinha}(2011)}]{Sadhu2011}%
  \BibitemOpen
  \bibfield  {author} {\bibinfo {author} {\bibfnamefont {J.}~\bibnamefont
  {Sadhu}}\ and\ \bibinfo {author} {\bibfnamefont {S.}~\bibnamefont {Sinha}},\
  }\href {\doibase 10.1103/PhysRevB.84.115450} {\bibfield  {journal} {\bibinfo
  {journal} {Physical Review B: Condensed Matter and Materials Physics}\
  }\textbf {\bibinfo {volume} {84}},\ \bibinfo {pages} {115450} (\bibinfo
  {year} {2011})}\BibitemShut {NoStop}%
\bibitem [{\citenamefont {Lim}\ \emph {et~al.}(2012)\citenamefont {Lim},
  \citenamefont {Hippalgaonkar}, \citenamefont {Andrews}, \citenamefont
  {Majumdar},\ and\ \citenamefont {Yang}}]{Lim2012}%
  \BibitemOpen
  \bibfield  {author} {\bibinfo {author} {\bibfnamefont {J.}~\bibnamefont
  {Lim}}, \bibinfo {author} {\bibfnamefont {K.}~\bibnamefont {Hippalgaonkar}},
  \bibinfo {author} {\bibfnamefont {S.~C.}\ \bibnamefont {Andrews}}, \bibinfo
  {author} {\bibfnamefont {A.}~\bibnamefont {Majumdar}}, \ and\ \bibinfo
  {author} {\bibfnamefont {P.}~\bibnamefont {Yang}},\ }\href {\doibase
  10.1021/nl3005868} {\bibfield  {journal} {\bibinfo  {journal} {Nano Letters}\
  }\textbf {\bibinfo {volume} {12}},\ \bibinfo {pages} {2475} (\bibinfo {year}
  {2012})}\BibitemShut {NoStop}%
\bibitem [{\citenamefont {Yang}\ \emph {et~al.}(2010)\citenamefont {Yang},
  \citenamefont {Zhang},\ and\ \citenamefont {Li}}]{Yang2010}%
  \BibitemOpen
  \bibfield  {author} {\bibinfo {author} {\bibfnamefont {N.}~\bibnamefont
  {Yang}}, \bibinfo {author} {\bibfnamefont {G.}~\bibnamefont {Zhang}}, \ and\
  \bibinfo {author} {\bibfnamefont {B.}~\bibnamefont {Li}},\ }\href {\doibase
  10.1016/j.nantod.2010.02.002} {\bibfield  {journal} {\bibinfo  {journal}
  {Nano Today}\ }\textbf {\bibinfo {volume} {5}},\ \bibinfo {pages} {85}
  (\bibinfo {year} {2010})},\ \Eprint {http://arxiv.org/abs/1002.3419v1}
  {1002.3419v1} \BibitemShut {NoStop}%
\bibitem [{\citenamefont {Maire}\ \emph {et~al.}(2017)\citenamefont {Maire},
  \citenamefont {Anufriev},\ and\ \citenamefont {Nomura}}]{Maire_2017}%
  \BibitemOpen
  \bibfield  {author} {\bibinfo {author} {\bibfnamefont {J.}~\bibnamefont
  {Maire}}, \bibinfo {author} {\bibfnamefont {R.}~\bibnamefont {Anufriev}}, \
  and\ \bibinfo {author} {\bibfnamefont {M.}~\bibnamefont {Nomura}},\ }\href
  {\doibase 10.1038/srep41794} {\bibfield  {journal} {\bibinfo  {journal}
  {Scientific Reports}\ }\textbf {\bibinfo {volume} {7}},\ \bibinfo {pages}
  {41794} (\bibinfo {year} {2017})}\BibitemShut {NoStop}%
\bibitem [{\citenamefont {Asmar}(2005)}]{Asmar2005}%
  \BibitemOpen
  \bibfield  {author} {\bibinfo {author} {\bibfnamefont {N.~H.}\ \bibnamefont
  {Asmar}},\ }\href@noop {} {\emph {\bibinfo {title} {Partial differential
  equations with Fourier series and boundary value problems}}}\ (\bibinfo
  {publisher} {Dover Publications},\ \bibinfo {address} {Upper Saddle River,
  N.J.},\ \bibinfo {year} {2005})\BibitemShut {NoStop}%
\bibitem [{\citenamefont {Dhar}(2001)}]{Dhar2001}%
  \BibitemOpen
  \bibfield  {author} {\bibinfo {author} {\bibfnamefont {A.}~\bibnamefont
  {Dhar}},\ }\href {\doibase 10.1103/PhysRevLett.86.5882} {\bibfield  {journal}
  {\bibinfo  {journal} {Physical Review Letters}\ }\textbf {\bibinfo {volume}
  {86}},\ \bibinfo {pages} {5882} (\bibinfo {year} {2001})}\BibitemShut
  {NoStop}%
\bibitem [{\citenamefont {Hou}\ \emph {et~al.}(2016)\citenamefont {Hou},
  \citenamefont {Xu}, \citenamefont {Ge},\ and\ \citenamefont {Li}}]{Hou2016}%
  \BibitemOpen
  \bibfield  {author} {\bibinfo {author} {\bibfnamefont {C.}~\bibnamefont
  {Hou}}, \bibinfo {author} {\bibfnamefont {J.}~\bibnamefont {Xu}}, \bibinfo
  {author} {\bibfnamefont {W.}~\bibnamefont {Ge}}, \ and\ \bibinfo {author}
  {\bibfnamefont {J.}~\bibnamefont {Li}},\ }\href {\doibase
  10.1088/0965-0393/24/4/045005} {\bibfield  {journal} {\bibinfo  {journal}
  {Modelling and Simulation in Materials Science and Engineering}\ }\textbf
  {\bibinfo {volume} {24}},\ \bibinfo {pages} {045005} (\bibinfo {year}
  {2016})}\BibitemShut {NoStop}%
\end{thebibliography}%

\end{document}